\documentclass[trackchanges, twocolumn]{aastex701}

\usepackage{enumerate, amsmath, longtable}

\begin{document}


\title{Chemical Signatures of AGB Mass Transfer in Gaia White Dwarf Companions}

\author[0000-0001-6970-1014]{Natsuko Yamaguchi}
\affiliation{Department of Astronomy, California Institute of Technology, 1200 E. California Blvd, Pasadena, CA, 91125, USA}
\email[show]{nyamaguc@caltech.edu}

\author[0000-0002-6871-1752]{Kareem El-Badry}
\affiliation{Department of Astronomy, California Institute of Technology, 1200 E. California Blvd, Pasadena, CA, 91125, USA}
\email{kelbadry@caltech.edu}

\author[0000-0001-6533-6179]{Henrique Reggiani}
\affiliation{Gemini South, Gemini Observatory, NSF’s NOIRLab, Casilla 603, La Serena, Chile}
\email{henrique.reggiani@noirlab.edu}

\author[0000-0001-8006-6365]{René Andrae}
\affiliation{Max-Planck-Institut für Astronomie, Königstuhl 17, D-69117 Heidelberg, Germany}
\email{andrae@mpia.de}

\author[0000-0001-9298-8068]{Sahar Shahaf}
\affiliation{Max-Planck-Institut für Astronomie, Königstuhl 17, D-69117 Heidelberg, Germany}
\email{sashahaf@mpia.de}


\begin{abstract}
We present a homogeneous abundance analysis of 160 main-sequence stars in astrometric white-dwarf + main-sequence (WD+MS) binaries with orbits from Gaia DR3. These systems have AU-scale separations and are thought to have undergone mass transfer (MT) when the WD progenitor was an asymptotic giant branch (AGB) star.  Using high-resolution spectroscopy, we measure chemical abundances of the MS stars, focusing on s-process elements. Since s-process nucleosynthesis occurs mainly in AGB stars, s-process enhancement in the MS star is a key signature of accretion from an AGB companion. We identify 43 barium dwarfs -- 39 of them newly discovered -- roughly doubling the known population in astrometric WD+MS binaries and extending it to lower metallicities than previously studied. The s-process abundances show large star-to-star variations that correlate with component masses and with metallicity but not with orbital separation. At the lowest metallicities, three barium dwarfs display strong CH and $\rm C_2$ absorption bands, confirming a link between barium stars and CEMP-s stars and implying that AGB mass transfer usually leads to strong carbon enhancement at low metallicity. By comparing the observed abundance patterns to AGB nucleosynthesis models, we show that the diversity of s-process enhancements can be explained by variations in donor mass, metallicity, and most importantly, the number of thermal pulses the AGB star experienced before the onset of MT. Variation in the depth of the accretors' convective envelopes, with which accreted material is diluted, strengthens correlations with MS star mass and metallicity. Our results establish Gaia WD+MS binaries -- which are homogeneously selected and probe shorter orbital periods than previous barium-star samples -- as a powerful laboratory for constraining mass transfer physics and the origin of chemically peculiar stars. \end{abstract}




\section{Introduction} \label{sec:Intro}

Binaries hosting white dwarfs (WDs) and main sequence (MS) stars in close orbits (separations $a \lesssim10\,$AU) are end products of mass transfer (MT) processes that occurred when the WD progenitor was a giant. This makes WD+MS binaries observational probes of the physics of MT, which remains one of the most important unsolved problems in stellar physics and binary evolution. 

The third data release (DR3) of the Gaia mission contained astrometric orbital solutions for over 160,000 binaries \citep{GaiaCollaboration2023, GaiaCollaboration2023A&A}. These solutions were obtained by fitting a Keplerian model to the wobble of unresolved binaries on the plane of the sky due to orbital motion. In the case where the secondary is an optically dark compact object such as a cool WD, the wobble is that of the primary star. Long-period orbits produce larger astrometric signals than short-period orbits, but orbits with periods significantly longer than the observational baseline ($\sim1000\,$d in DR3) cannot be constrained astrometrically. As a result, DR3 astrometric orbits are sensitive to orbital periods ($P_{\rm orb}$) ranging from roughly 100 to 1000 days. Using the DR3 astrometric orbit catalog, \citet{Shahaf2024MNRAS} identified thousands of new WD+MS binary candidates. Most of them host $\sim 0.6\,M_{\odot}$ WDs in AU-scale orbits, suggesting interaction with an asymptotic giant branch (AGB) donor \citep{Hallakoun2024ApJL}. While this is classically expected to lead to unstable mass transfer, their orbits are significantly wider than previously identified post-common envelope binaries (PCEBs) with $P_{\rm orb} \lesssim 1\,$d \citep[e.g.][]{Zorotovic2010A&A, Zorotovic2011A&A, Parsons2015MNRAS}. Modeling of the selection function reveals that these AU-scale systems are common but were disfavored by most discovery methods before Gaia \citep{Yamaguchi2024PASP_SLBs, Yamaguchi2025arXiv}.

Several recent works have explored possible formation scenarios for AU-scale WD+MS binaries, finding that (1) MT from AGB donors may remain stable over a wider range of initial mass ratios than previously thought (based on ``quasi-adiabatic" critical mass ratios derived by e.g. \citealt{Ge2020ApJ, Temmink2023A&A} in the general case of evolved donors), and (2) PCEBs may remain wide if MT begins during a thermal pulse of an AGB donor, when its envelope is very loosely bound (e.g. \citealt{Belloni2024, Yamaguchi2024PASP, Yamaguchi2024MNRAS}, who studied the energy budget of the resulting common envelope in the context of the Gaia systems). Still, a formation model that can explain all of the distinctive traits of the Gaia post-MT systems (intermediate orbital separations, non-zero eccentricities, and a narrow WD mass distribution), as well as the sheer abundance of such systems, remains elusive.

One key observational signature of a previous MT phase is the presence of anomalous surface abundances. In particular, barium stars are stars with enhanced abundances of s-process elements, including but not limited to barium. Since the first studied barium stars were giants, the term ``barium dwarfs" is commonly used for MS stars enhanced in s-process elements. As s-process nucleosynthesis is expected only to occur during the late stages of stellar evolution on the AGB, the abundances of these pre-AGB stars have been attributed to s-process enhanced material accreted from an AGB companion. Indeed, since their discovery, many barium stars have been confirmed to be in binary systems with WD companions \citep[e.g.][]{McClure1980ApJL, McClure1983ApJ, Jorissen1988A&A}. 

Another closely related population is the carbon enhanced metal-poor (CEMP) stars found in the Galatic halo, with typical abundances of $\mathrm{[Fe/H]}\lesssim-2.0$ and $\mathrm{[C/Fe]}\gtrsim0.7$ \citep[e.g.][]{Rossi1999ASPC, Beers2005ARA&A, Aoki:2007ApJ...655..492A}. A large fraction of CEMP stars are also s-process enhanced, with $\mathrm{[Ba/Fe]}\gtrsim0.5$ (the so-called `CEMP-s' stars; e.g. \citealt{Norris1997ApJ, Aoki2007ApJ, Jonsell2006A&A, Masseron2010A&A}). Similar to most barium stars, the majority of CEMP-s stars reside in binaries \citep{Lucatello2005ApJ} and thus, their s-process enhancement has also been attributed to mass transfer from AGB donors \citep{Abate2015A&A, Abate2018A&A}. However, compared to barium stars, CEMP-s stars are more pristine, meaning it is easier to disentangle signatures of metals originally present in their atmosphere and those of accreted material. They are also older, making them unique probes of binary interactions in the early Universe \citep[e.g.][]{Arentsen2022MNRAS}. Additionally, dwarf carbon (dC) stars share similar properties to CEMP stars, but as they are less evolved, the origin their carbon enhancement is more narrowly restricted to accretion from a companion \citep{Dearborn1986, Green1991ApJL, Green2013ApJ, Downes2004AJ, Farihi2018MNRAS, Farihi2025MNRAS, Roulston2019ApJ}.

Recently, \citet{Rekhi2024ApJL, Rekhi2025arXiv} compiled a sample of 38 systems with s-process enrichment in the Gaia sample of WD+MS binaries using a combination of GALAH archival spectra and their own follow-up spectra. They found that about 30\% of WD+MS binaries in the Gaia sample are enhanced in s-process elements, but the s-process abundances are not strongly correlated with the binaries' orbital parameters. They also identified an anti-correlation between s-process abundances and bulk metallicity in the sample. 

In this paper, we expand on these works by targeting a larger sample of WD+MS binaries from the Gaia sample with high-resolution spectroscopy. We target sources with a broad range of metallicities, including several below $\rm [Fe/H] = -2$, and also include several candidate MS + neutron star (NS) binaries. We identify 39 new barium dwarfs, and confirm 4 existing ones in the literature. We find evidence of significant carbon enhancement in three of the most metal-poor barium dwarfs, creating a link to CEMP-s stars. We also present a population model to explain the diversity of s-process abundances found in the Gaia WD+MS binary sample.

A considerable body of previous work has studied the population demographics of  chemically peculiar stars (e.g. \citealt{Escorza2019A&A, Abate2015A&A, Abate2018A&A, Hansen2016A&A_2, Dimoff2025A&A}). These works have confirmed that accretion from an AGB companion can broadly explain the observed abundances and have established evolutionary links between several related classes of chemically peculiar systems. However, previous studies have included only approximate treatments of sample selection functions, making detailed comparisons between models and data challenging.  Moreover, many of the previously identified systems have orbital periods longer than $1000\,$d, where wind accretion or wind Roche lobe overflow likely dominate the MT process. Our work thus adds to the existing literature by focusing on systems with orbital solutions from Gaia that have $100-1000\,$d periods. This means that they have a homogeneous and well-understood selection function, and their AU-scale orbits suggest that they likely experienced traditional Roche lobe overflow.

The remainder of this paper is organized as follows. In Section \ref{sec:selection}, we describe target selection for follow-up optical spectroscopy. In Section \ref{sec:observations}, we describe observations of these targets and the data reduction process. Section \ref{ref:analysis} details the analysis to constrain atmospheric parameters and measure chemical abundances. In Section \ref{sec:results}, we discuss the relationships between the measured abundances and how they vary with the various properties of the systems. We also describe two barium dwarfs orbiting candidate NSs and three with evidence of strong carbon enhancement. In Section \ref{sec:literature}, we compare our barium dwarfs to several related populations of chemically peculiar stars in the literature. In Section \ref{sec:theory}, we use theoretical models of s-process yields in the AGB donors and consider the effect of convective mixing in the envelopes of the MS accretors to predict and explain the observed trends and scatter in the barium abundances with the stellar and orbital parameters. Finally, we conclude with a summary of our key findings in Section \ref{sec:conclusion}.

\section{Target selection} \label{sec:selection}

\begin{figure*}
    \centering
    \includegraphics[width=0.9\linewidth]{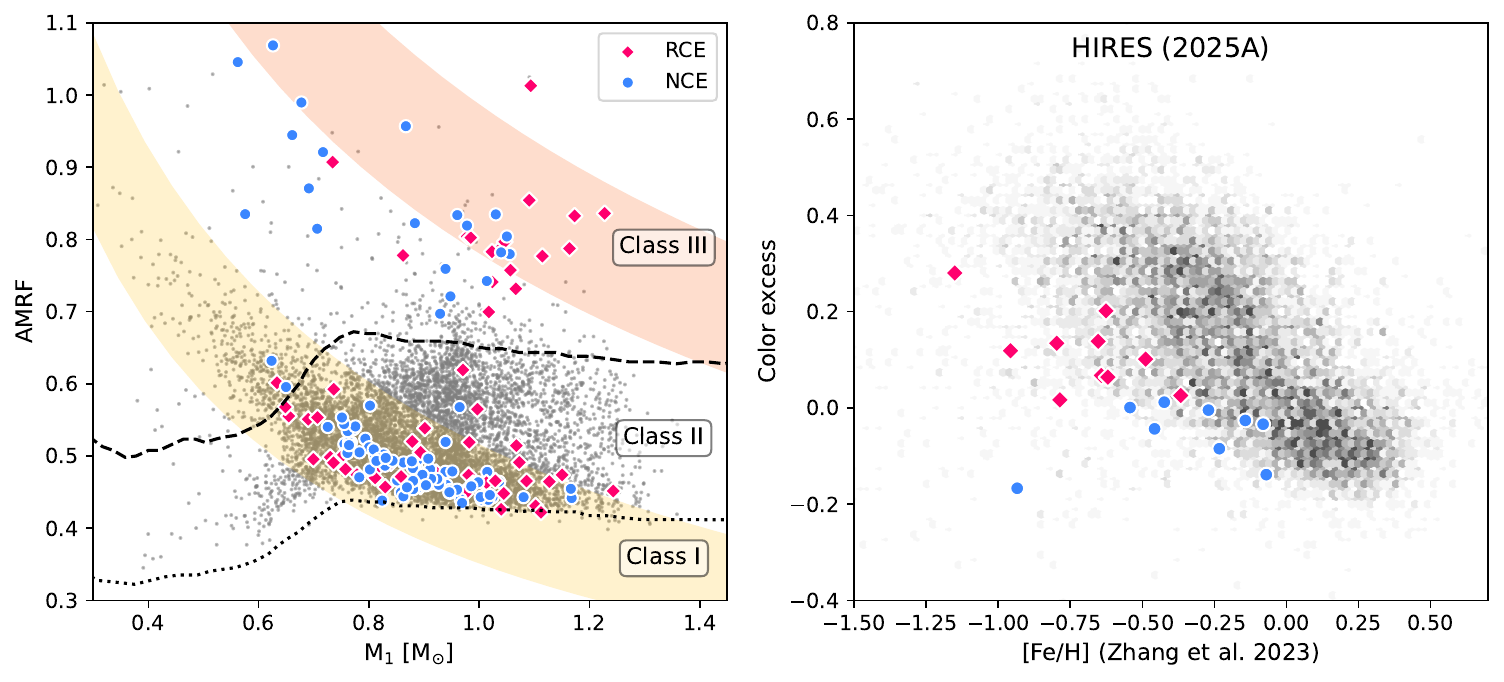}
    \caption{\textit{Left}: AMRF versus luminous star mass, $M_1$, for the full non-class I sample from \citet{Shahaf2024MNRAS} (grey points). Our targets selected for follow-up are shown with colored points, separated by whether they were classified as NCE (pink diamonds) or RCE (blue circles) by \citet{Shahaf2024MNRAS}. The dotted and dashed lines are boundaries separating systems in the three classes (Section \ref{sec:selection}). The shaded regions corresponds to the expected locations of systems hosting dark compact object secondaries with masses between $0.45-0.75\,M_{\odot}$ (yellow, lower) and $1.4-2.1\,M_{\odot}$ (orange, upper), corresponding to typical WD and NS masses. While sources classified as NCE are less likely to be triples, we believe that both the NCE and RCE sources in our follow-up sample are dominated by compact object companions (see text). \textit{Right}: Color excess against [Fe/H] \citep{Zhang2023MNRAS} for non-class I systems. The two quantities are anti-correlated, meaning that the NCE sample is biased against metal-poor primaries. We selected several RCE systems lying below the densely populated strip in this space, which are likely to host WDs.}
    \label{fig:selection}
\end{figure*}

We selected targets from the sample of  sources identified by \citet{Shahaf2024MNRAS} as unlikely to be MS+MS binaries. Their selection relies on a quantity called the astrometric mass ratio function (AMRF), which can be defined purely in terms of observables:
\begin{equation}  \label{eqn:AMRF_1}
    \mathrm{AMRF} = \frac{\alpha}{\varpi}\left( \frac{M_1}{M_{\odot}} \right)^{-1/3}\left( \frac{P_{\rm orb}}{\mathrm{yr}} \right)^{-2/3}
\end{equation}
where $\alpha$ is the angular photocentric semi-major axis, $\varpi$ is the parallax, $M_1$ is the mass of the luminous star, and $P_{\rm orb}$ is the orbital period. As described by \citet{Shahaf2019MNRAS}, observed binaries can be grouped into three classes based on their location in the AMRF vs $M_1$ plane.
\begin{itemize}
    \item Class-I: Orbit consistent with hosting a single MS secondary. 
    \item Class-II: Orbit inconsistent with hosting a single MS secondary. May host an inner binary of two low-mass MS stars (forming a triple system), or a compact object contributing little to no light (most likely a cool WD, but also possibly a NS or black hole).  
    \item Class-III: Orbit inconsistent with hosting a single MS companion or an inner binary containing two MS stars. Likely to host a compact object. 
\end{itemize}

In our spectroscopic follow-up, we considered both class-II and III  (i.e. ``non-class-I") systems. In the left panel of Figure \ref{fig:selection}, we show our targets in the AMRF$-M_1$ plane. While restricting to only class-III systems would provide a purer sample of compact objects with less potential contamination from triples, this would also exclude the vast majority of systems hosting WDs, which have mass ratios too low to be class-III. Instead, in constructing their WD + MS binary catalog, \citet{Shahaf2024MNRAS} removed potential triples in the class-II systems by implementing a color cut which excludes systems that appear redder than expected for a single MS star. They call the resulting sample the ``No Color Excess (NCE)" sample. The remaining non-class I systems that do not pass this cut -- meaning they are redder than predicted by a 2 Gyr-old isochrone at the same metallicity -- are referred to as the ``Red Color Excess (RCE)" sample.

We prioritized NCE candidates in our follow-up. However, many of our targets were selected from a preliminary, unpublished version of the \citet{Shahaf2024MNRAS} sample, and were classified as NCE there but as RCE in the final sample. The preliminary and final samples differed in the adopted metallicities and color excess thresholds. We found that the final NCE selection of \citet{Shahaf2024MNRAS} was conservative: most of the systems classified as RCE class-II in our sample are still likely to host  WDs, as evidenced by their mass and eccentricity distributions. Moreover, the fact that many of our class-III targets -- which have AMRF values too large to be explained by triples -- are nevertheless classified as RCE provides further evidence that many of the objects labeled as RCE are not actually triples.

Simulations from \citet{Yamaguchi2025arXiv} showed that while the cut on color excess effectively eliminates triples, it also removes about half of all systems hosting WDs in the non-class I sample (see their Figure 5). In particular, they found that the cut is biased against metal-poor systems, likely because the adopted 2 Gyr-old isochrones predict a relation between color and metallicity that is steeper than observed. This means that very few systems with [Fe/H] $< -0.5\,$dex enter the NCE sample, suggesting that a metallicity-dependent cut on the color excess may be more appropriate than the flat cut adopted by \citet{Shahaf2024MNRAS}.

As we discuss in Section \ref{ssec:cemp_stars}, the most metal-poor barium dwarfs also tend to be enhanced in carbon. Because carbon forms strong molecular bands in the blue optical (Figure \ref{fig:ch_g_band}), stars that are carbon-rich will appear redder than those that are not \citep[e.g.][]{Ardern-Arentsen2025MNRAS}. While the magnitude of the color excess resulting from this effect will depend on the exact carbon abundance, this is another caveat of the NCE sample.

With these effects in mind, we selected several metal-poor systems to be observed with HIRES (Section \ref{sec:observations}) that are in the non-class I sample, but not in the NCE sample of \citet{Shahaf2024MNRAS}. As shown in the right panel of Figure \ref{fig:selection}, we selected metal-poor stars that lie below the average color excess at their metallicties. 

In total, we observed 178 sources. 

\section{Observations} \label{sec:observations}

We obtained high-resolution spectra of targets through several different programs using three instruments over the course of two years, as described below.

\subsection{FEROS}

We obtained a total of 219 spectra of 92 targets using the Fiber-fed Extended Range Optical Spectrograph (FEROS; \citealt{Kaufer1999Msngr}) instrument installed on the MPG/ESO $2.2\,$m telescope at the La Silla Observatory (Programs 113.26XB, 114.27SS, and 115.28KE). Exposure times ranged from 600 to 2400 s, depending on the brightness of the target. We used $1\times1$ binning. This allowed us to achieve a spectral resolution $R\sim50,000$ over $3860$ to $6770\,$\AA. For a single observation, the typical signal-to-noise ratio (SNR) achieved was $\sim25$ and $15$ above and below $5000\,$\AA, respectively. We reduced the raw data using the CERES pipeline \citep{Brahm2017PASP}. 

\subsection{MIKE}

We obtained 54 spectra of 24 targets with the Magellan Inamori Kyocera Echelle (MIKE) spectrograph mounted on the Magellan Clay telescope at Las Campanas Observatory \citep{Bernstein2003SPIE}. We used the $0.7''$ slit with $2\times2$ binning, which resulted in a typical $R\sim34,000$ and wavelength coverage of $3325-9680\,$\AA. Exposure times ranged from 600 to 2400$\,$s, which yielded typical SNR for a single observation of $\sim50$ and $40$ above and below $5000\,$\AA. We reduced the spectra with the MIKE Pipeline using CarPy \citep{Kelson2000ApJ, Kelson2003PASP}. 

\subsection{HIRES}

We obtained 66 spectra of 61 targets with the High Resolution Echelle Spectrometer (HIRES) on the Keck I telescope. 34 of these targets were observed through the California Planet Search (CPS; \citealt{Howard2010ApJ}) queue, which allowed for more flexible scheduling. For all observations, we used the C2 decker ($0.86''\times 14''$ slit) with $3\times1$ binning, yielding spectra with $R\sim50,000$. For targets observed through CPS, our exposure times ranged from $120$ to $1200\,$s and we used the HIRESr cross-disperser. This resulted in SNR $\sim55$ and $35$ above and below $5000\,$\AA \ and wavelength coverage of $3640-6420\,$\AA. For the other targets, we used a fixed exposure time of 1200$\,$s and the HIRESb cross-disperser. This led to SNR $\sim40$ and $35$ above and below $5000\,$\AA, and wavelength coverage $3840-6640\,$\AA. 

We reduced the raw data using the MAuna Kea Echelle Extraction (MAKEE) package, which extracts spectra from each echelle order, then carries out bias correction, background subtraction, and wavelength calibration. 

We note that a subset of the MIKE and HIRES observations were originally made as part of a follow-up program of NS + MS binary candidates \citep{El-Badry2024OJAp_NS, El-Badry2024OJAp_NSpop}, but were primarily used to measure radial velocities (RVs) rather than abundances.

\subsection{Continuum normalization and RV correction}

We processed all reduced spectra using the \texttt{iSpec} package \citep{Blanco-Cuaresma2014A&A, Blanco-Cuaresma2019MNRAS}. To estimate the continuum for each echelle order, \texttt{iSpec} first applies a median and maximum filter to the data to mitigate noise effects and avoid strong absorption features, and then fits multiple splines of a specified degree to the data in each order to be used for the normalization. We chose typical window sizes of $0.05\,$nm and $1.0\,$nm for the median and maximum filters, respectively, and splines of degrees 4 and 5, but varied the exact choices for each order and instrument in order to optimize the continuum normalization and avoid overfitting \footnote{For more details of this process, consult the \texttt{iSpec} manual: www.blancocuaresma.com/s/iSpec/manual/introduction}. We then merged all orders of the normalized spectra, taking an inverse-variance weighted average of the flux in overlap regions. The RV of each spectrum was calculated with \texttt{iSpec}'s cross-correlation algorithm using a template solar spectrum. We provide a table of RVs in Appendix \ref{appendix:rv_tables}. After shifting the spectra to rest frame, if repeated observations were taken for one target with the same instrument, we co-added them into one spectrum to increase SNR. 

\section{Analysis} \label{ref:analysis}

\begin{figure*}
    \centering
    \includegraphics[width=0.95\linewidth]{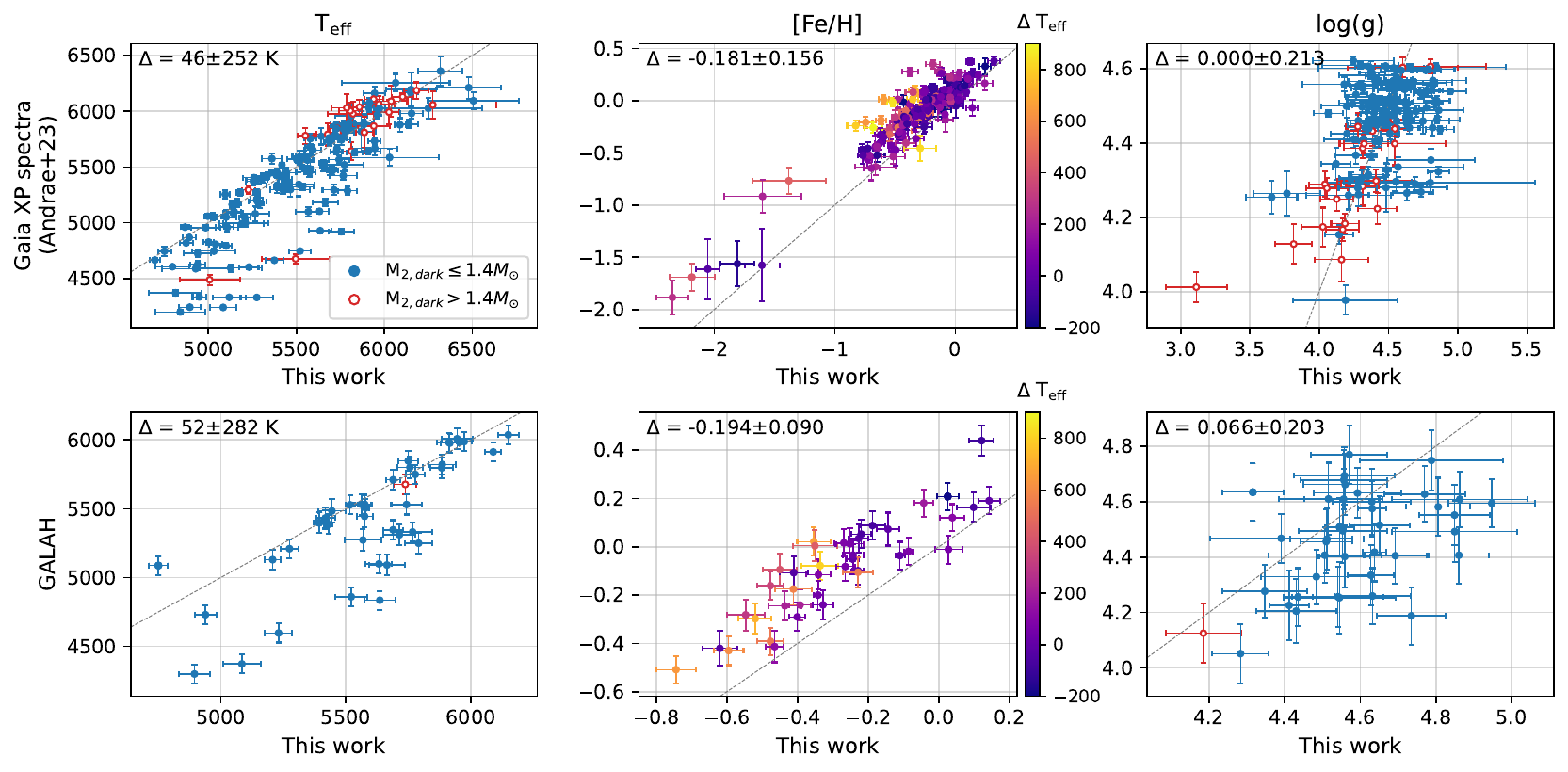}
    \caption{Comparison of best-fit stellar parameters obtained in this work (Section \ref{ssec:sp_ew_balance}) with those calculated using Gaia low-resolution spectra (\citealt{Andrae2023ApJS}; top row), as well as those from GALAH DR4 (\citealt{Buder2025PASA}; bottom row). In the upper left of each panel, we note the median and standard deviation of the offsets ($\Delta$ = This work - Gaia XP/GALAH). For plots comparing [Fe/H], we color the points by the value of the offsets in $T_{\rm eff}$. For the remaining panels, we show systems with secondary masses above and below $1.4\,M_{\odot}$ with red unfilled and blue filled markers, respectively. }
    \label{fig:sp_comparison}
\end{figure*}

\subsection{Atmospheric parameters with excitation/ionization balance} \label{ssec:sp_ew_balance}

To measure equivalent widths (EWs) of absorption lines, we used \texttt{REvIEW}\footnote{https://github.com/madeleine-mckenzie/REvIEW} \citep{McKenzie2022MNRAS}. Provided a spectrum and linelist, \texttt{REvIEW} fits a region $\pm0.6\,$\AA \ around the central wavelength of each line with up to three Gaussians to separate out contributions from closely spaced lines that may be partly blended. 

We then use \texttt{q2}\footnote{https://github.com/astroChasqui/q2} \citep{Ramirez2014} to determine stellar parameters. This code is a python interface for \texttt{MOOG} (2019 release; \citealt{Sneden1973PhDT, Sneden2012ascl}) and calls the \texttt{abfind} function, which uses the excitation/ionization equilibrium method with the measured EWs of iron lines. This method determines the optimal set of effective temperature ($T_{\rm eff}$), surface gravity (log$(g)$), metallicity ([Fe/H]), and microturbulence ($\xi$) for which the inferred absolute iron abundance ($A$(Fe))) from individual lines is independent of the lines' ionization states, excitation potentials (EP), and relative line strengths (i.e. reduced equivalent width, REW) \citep[e.g.][]{Gray2008oasp}. \texttt{q2} calculates errors on the stellar parameters by computing the effect of small perturbations in these parameters to the equilibrium conditions, following the methods of \citet{Epstein2010ApJ} and \citet{Bensby2014A&A}.

We remove lines with EW above $120\,$m\AA \ to avoid saturated lines, as well as those with EW below $10\,$m\AA, which we found to be dominated by noise. In cases with low SNR, we increased this lower bound to $20\,$m\AA. We use the MARCS atmosphere grid \citep{Gustafsson2008A&A} and by default, we fit all four stellar parameters. For the initial guess of $T_{\rm eff}$, log$(g)$, and [Fe/H], we used best-fit values obtained using the {\it Gaia} XP very low-resolution spectra, as described in \citet{Andrae2023ApJS}. However, we do not use published results from \citet{Andrae2023ApJS} as they make use of parallaxes from the \texttt{gaia\_source} catalog, which are derived from a single-star astrometric solution. Instead, we use re-calculated values using parallaxes from the \texttt{nss\_two\_body\_orbit} table (Appendix \ref{appendix:xgboost}). For $\xi$, we use the empirical relation from \citet{Ramirez2013ApJ} (their Equation 5), derived for FGK MS stars.

For the majority of spectra (131 out of 178), the method described above leads to successful convergence of stellar parameters. In Appendix \ref{appendix:q2_plots}, we show the excitation/ionization balance achieved for several objects with both successful and problematic results. For successful cases, the inferred abundances depend minimally on line-specific properties. 

The remaining spectra experience one or more of the following issues: 
\begin{enumerate}[i.]
    \item Best-fit $\xi$ goes to 0.
    \item Best-fit log$(g)$ goes to 5.0.
    \item Convergence of stellar parameters is not achieved.
\end{enumerate}
The first two issues are problematic as solvers are reaching the edge of the stellar atmosphere grid, making the results potentially unreliable. In addition, based on the position of our objects on the color-magnitude diagram (CMD), all primaries are expected to be MS stars with log$(g) = 4-5$. These issues may be traced back to inaccurate EW measurements due to low-SNR spectra, spectra broadened by stellar rotation, and/or poor choice of parameter initialization. Moreover, some stars have few strong Fe II lines with reliable EWs, which can prevent a robust constraint on log$(g)$. Additionally, $\xi$ is known to be sensitive to various systematics -- such as the SNR of the spectra, continuum placement, and line blending \citep[e.g.][]{Mucciarelli2011A&A, Jofre2014A&A, Blanco-Cuaresma2019MNRAS}. 

In these cases, we fix $\xi$ and/or log$(g)$ to their initial values and rerun the fit, leaving the remaining parameters free. The majority of objects for which fixing log$(g)$ was required were observed by FEROS, suggesting that the issue stems from their lower SNR compared to observations made with other instruments (Section \ref{sec:observations}). 

In total, we obtain reliable stellar parameters for 161 targets. The remaining 17 targets have spectra that did not achieve convergence or resulted in qualitatively poor excitation/ionization balance. Some of these were obtained in poor conditions, resulting in abnormally low SNR, prohibiting accurate EW measurements. This is particularly problematic for metal-poor stars with few strong iron lines. A few stars also have broad lines indicative of rapid rotation, which can cause blending and erroneous EWs. The rest of our analysis is focused on the 160 sources with converged and reliable parameters.

\subsection{Spectral fitting results} \label{ssec:spectral_fitting}

In the upper panels of Figure \ref{fig:sp_comparison}, we compare stellar parameters from our analysis to those calculated with the Gaia XP spectra using the method from \citet{Andrae2023ApJS}. We find general agreement between the two works. However, there are two notable discrepancies. Below $\sim 5500\,$K, we find a roughly linearly increasing discrepancy in  $T_{\rm eff}$, where our estimates are consistently hotter. This is not unexpected given previous work which has also found that stellar parameters determined using different methods and codes tend to become more discrepant with each other at low temperatures \citep[e.g.][]{Hegedus2023A&A}. We use the spectroscopic $T_{\rm eff}$ measurements throughout for consistency. In Appendix \ref{appendix:teff_err} (Figure \ref{fig:teff_test}), we show that our conclusions about which sources have s-process enhancement are not sensitive to this choice. 

In addition, our metallicities are systematically lower by 0.2 dex on average than those estimated by \citet{Andrae2023ApJS}. One factor that may contribute to this is that \citet{Andrae2023ApJS} provides the mean metallicity, [M/H], not [Fe/H], meaning that it is not a one-to-one comparison. In particular, old stars deficient in iron ([Fe/H]$<-1.0$) are usually more enriched in other metals, such as $\alpha$-elements \citep[e.g.][]{Cayrel2004A&A, Adibekyan2012A&A} and carbon \citep[e.g.][]{Li2022ApJ}, which would increase the average metal content. 

For objects found in the fourth data release of the Galactic Archaeology with HERMES (GALAH) Survey \citep{Buder2025PASA}, we also compare the quoted stellar parameters from this catalog (with quality flags \texttt{flag\_sp} = 0 and \texttt{flag\_fe\_h} = 0) to our results in the lower panels. We once again note an offset in [Fe/H]. We discuss how this offset affects our final barium dwarf fractions in Section \ref{ssec:GALAH_param_tests}.

While these comparisons are helpful for validation, we emphasize that \citet{Andrae2023ApJS}, GALAH, and our work all use distinct methods to estimate stellar parameters and metallicities (data-driven machine learning, synthetic spectral fitting with neural networks, and excitation/ionization balance, respectively) and different underlying radiative transfer codes, meaning systematic offsets are not unexpected \citep[e.g.][]{Blanco-Cuaresma2019MNRAS}. 

\begin{figure*}
    \centering
    \includegraphics[width=0.95\linewidth]{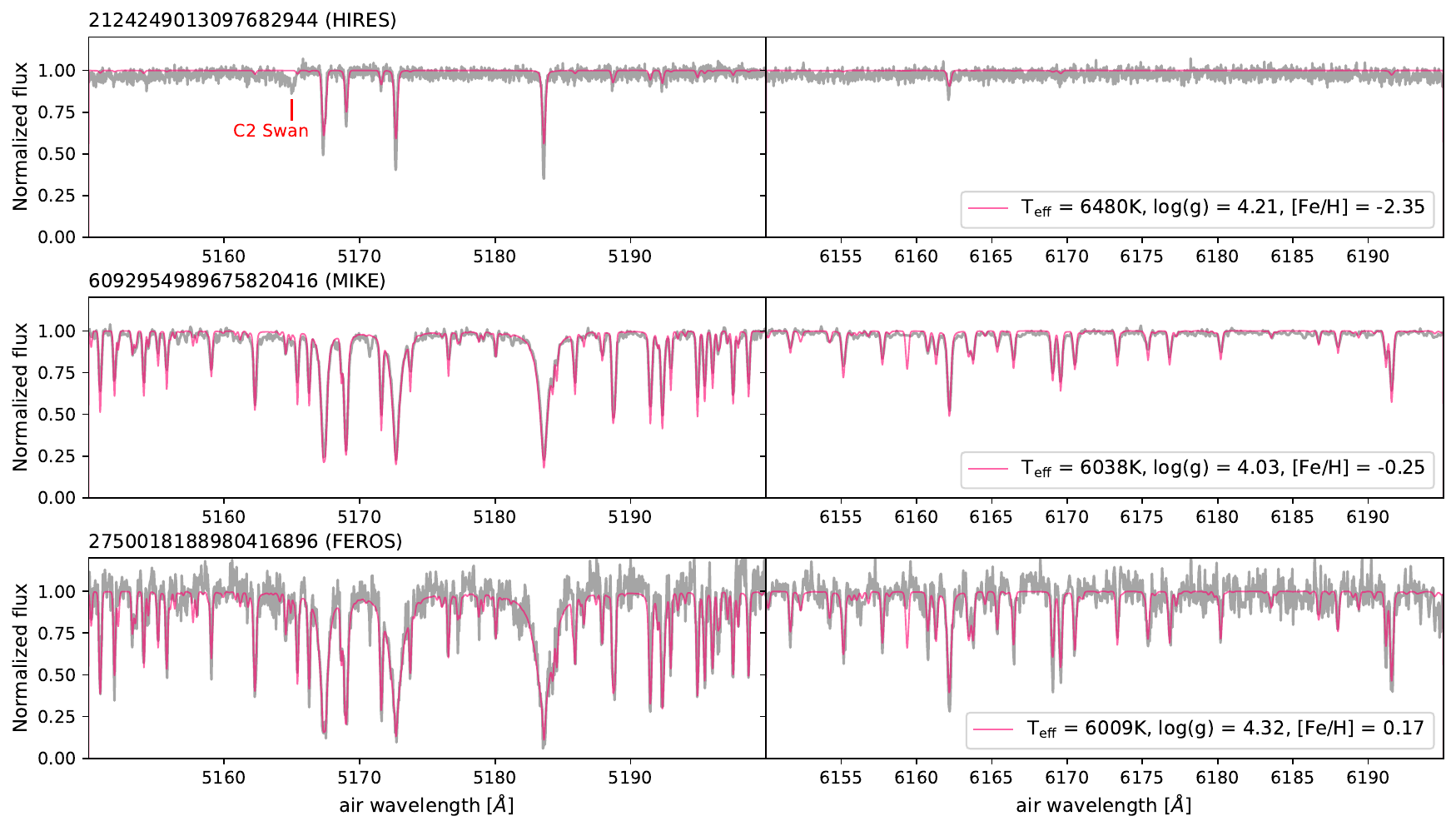}
    \caption{Observed spectra of three targets (gray) compared with synthetic spectra generated using their best-fit stellar parameters (pink; Section \ref{ssec:sp_ew_balance}). We select targets covering a wide range of [Fe/H] with similar $T_{\rm eff}$ and log$(g)$, observed with different instruments. On the left, we plot the region around the magnesium triplet, and on the right, we move to redder wavelengths populated with iron lines. Visually, we see good agreement between the observations and the corresponding models, suggesting that our stellar parameters are reasonable. For the most metal-poor star (top row), the C$_2$ Swan band at $5165\,$\AA \ is visible, which suggests carbon enhancement (Section \ref{ssec:cemp_stars}). }
    \label{fig:synspecs}
\end{figure*}

In Figure \ref{fig:synspecs}, we compare regions of the observed spectra for three objects to synthentic spectra generated using our best-fit $T_{\rm eff}$, log$(g)$, and [Fe/H]. We generated the synthetic spectra using the \texttt{iSpec} \citep{Blanco-Cuaresma2014A&A, Blanco-Cuaresma2019MNRAS}, which provides a python interface through which we employed the SPECTRUM radiative transfer code\footnote{https://www.appstate.edu/~grayro/spectrum/spectrum.html} \citep{Gray1994AJ}. The projected rotational velocity was fixed to the solar value ($v\sin(i)=1.6\,$km/s; e.g. \citealt{Pavlenko2012MNRAS}) and the resolution was set to 50,000. We did not vary individual elemental abundances when generating these models. We include these plots for objects observed with different instruments and at a range of metallicities to provide visual confirmation that the best-fit parameters obtained through our method described above are reasonable. 

\subsection{Deriving abundances} \label{ssec:abundances_derivation}

\begin{figure*}
    \centering
    \includegraphics[width=0.95\linewidth]{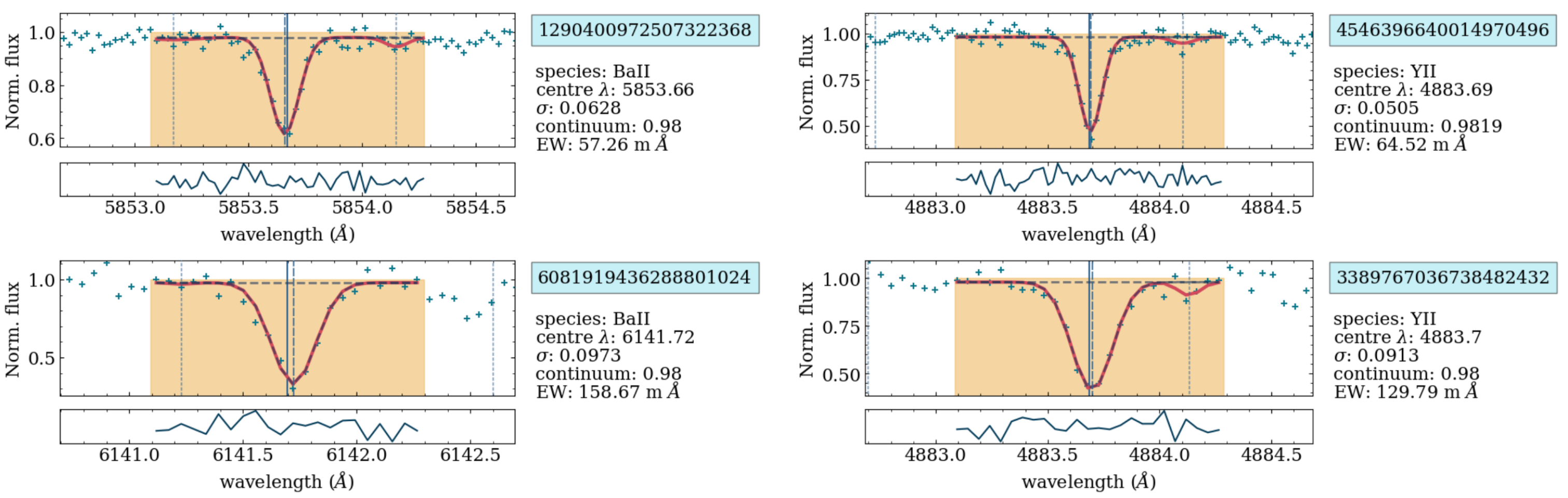}
    \caption{Examples of strong lines of singly ionized barium and yttrium. The fitting region is shaded in yellow, and the best-fit model is plotted in red. The solid vertical line indicates the position of the line center provided in the linelist, while the dashed and dotted lines indicate the minima of the fitted Gaussians. The upper two panels are fits to HIRES spectra, while the lower left and right are fits to FEROS and MIKE spectra.}
    \label{fig:line_fits}
\end{figure*}

To measure elemental abundances, we once again use \texttt{REvIEW} to measure EWs of lines for elements other than iron. In Figure \ref{fig:line_fits}, we show Gaussian fits to strong barium and yttrium lines in a few example barium dwarfs.  

Together with the best-fit stellar parameters determined above, the EWs are inputted into \texttt{q2}, which returns absolute abundances of specified elements inferred by each line. We consider lines with EWs between 20 and $180\,$m\AA. Compared to iron, we set a stricter lower bound to avoid the weakest lines where noise is being fit; while we set a less restrictive upper bound as there are fewer strong lines. Most elements of interest have at least one line with EW below this looser upper limit, and we have confirmed that the inferred abundances from individual lines with EWs above and below $120\,$m\AA \ are consistent, suggesting that saturation is not a major issue. In Table \ref{tab:strong_lines}, we list key lines of elements used to determine their abundances which are discussed in Section \ref{sec:results}. Note that we completely exclude several strong lines (e.g. Ba\,II $\lambda$4554\AA) as these are often saturated, making their EWs uninformative.

For each element, we discard lines which have residuals of more than 0.3 dex from the median. Unless otherwise stated, we use solar abundances from \citet{Asplund2021A&A} to calculate the differential abundances. Where there are multiple lines, the best-fit value is taken to be the median of the abundances inferred from each line and the error is taken to be their standard deviation divided by the square root of the total number of lines. If only one line is measured, we set a fixed error of 0.2 dex. This is a conservative choice, being roughly the 90th percentile of the standard deviations of abundances with multiple lines. 

We note that as expected, there is close agreement between [Fe\,I/H] and [Fe\,II/H] for most of our systems. However, approximately 17\% of them have discrepancies between greater than $0.2\,$dex. Almost all of these are cases where log$(g)$ was fixed to achieve convergence of stellar parameters (Section \ref{ssec:sp_ew_balance}). Since log$(g)$ is determined via ionization balance, this is unsurprising.

\begin{table}
\centering
\begin{tabular}{l|p{6cm}}
\hline
        Element & Central wavelengths [\AA] \\ \hline\hline
        Mg\,I & 4730.04, 5711.09 \\
        Ca\,I & 5867.56, 6499.65 \\
        Si\,I & 5645.61, 5665.55, 5684.48, 5690.42 5701.1, 5793.07, 6721.85 \\
        Ba\,II & 5853.67, 6496.9 \\
        Y\,II & 4854.87, 4883.69, 4900.11 \\
\hline
\end{tabular}
\caption{Lines of several elements of interest used to calculate abundances. All wavelengths are in air and rounded to two decimal places.}
\label{tab:strong_lines}
\end{table}

\section{Results} \label{sec:results}

The resulting best-fit stellar parameters and abundances for all 160 objects are summarized in Tables \ref{tab:stellar_params} and \ref{tab:abuns} in Appendix \ref{appendix:full_tables}. 

\subsection{Abundance patterns} \label{ssec:results_abundance_patterns}

\begin{figure*}
    \centering
    \includegraphics[width=0.95\linewidth]{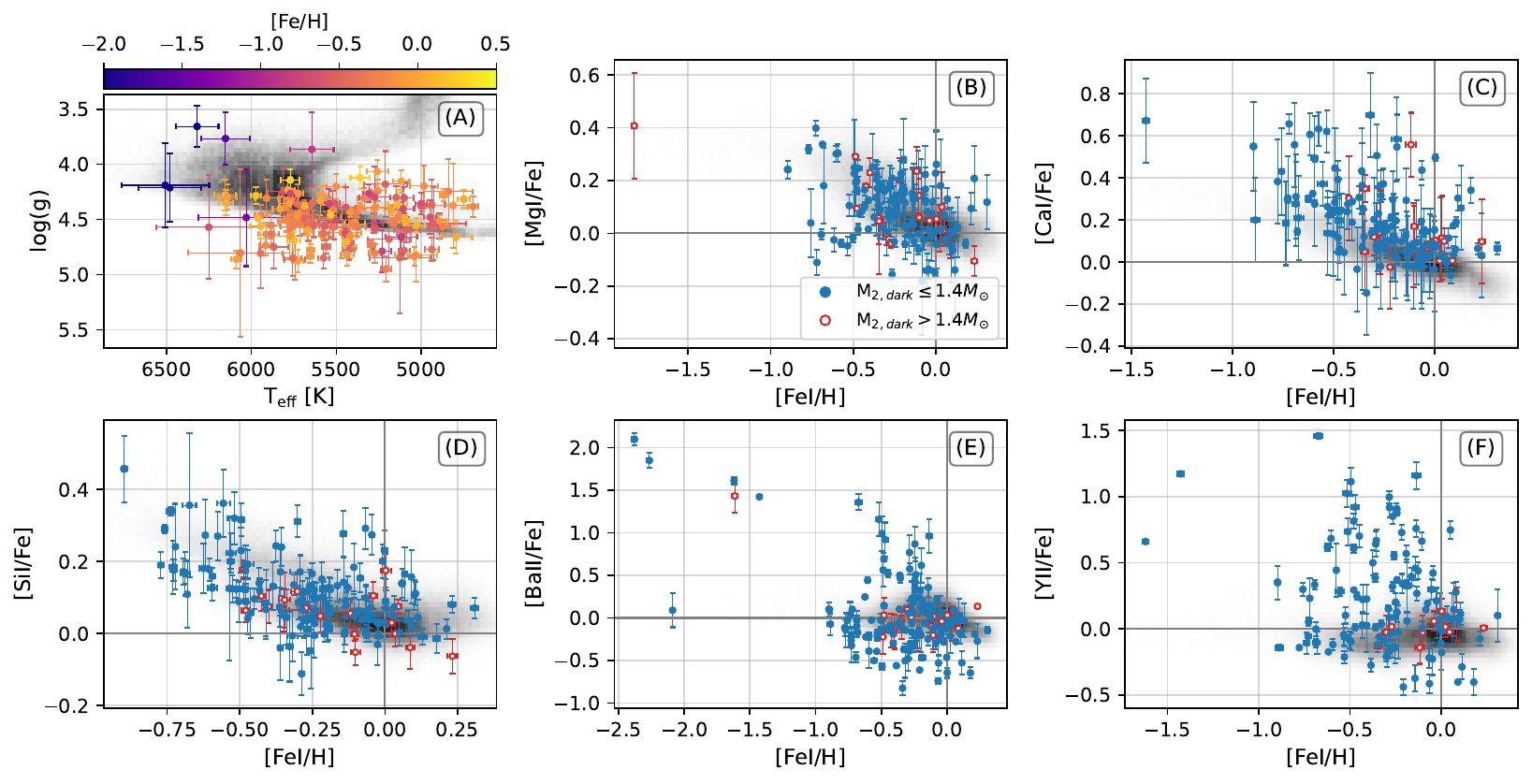}
    \caption{Atmospheric parameters (A) and derived abundances of various elements (B-F). In the background of each panel, we plot the distribution of the corresponding parameters for sources in the GALAH DR4 catalog. Panel (A) shows that as expected, all targets are consistent with being on the MS, with more metal-poor stars being hotter. Panels (B) - (D) show that the expected trends in the $\alpha$-elements are seen in both GALAH and our sample. We target stars that likely accreted material from an AGB donor, explaining the greater fraction of stars that are rich in s-process elements, seen in panels (E) and (F).}
    \label{fig:abundances}
\end{figure*}

\begin{figure*}
    \centering
    \includegraphics[width=0.95\linewidth]{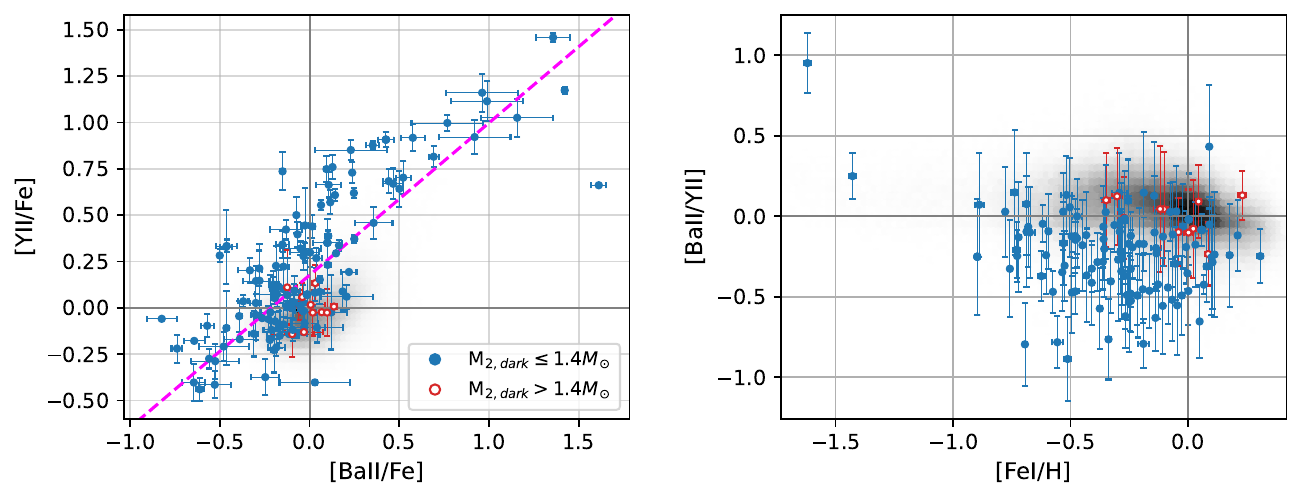}
    \caption{\textit{Left}: Abundance of yttrium plotted against barium. Both elements are primarily produced by the s-process in AGB stars. Their abundances are strongly correlated. Only sources with $M_{\rm 2, dark} < 1.4\,M_{\odot}$ are enhanced in both elements. The magenta line is a linear fit to the data. \textit{Right}: Relative abundance of barium (a heavy s-process element) to yttrium (a light s-process element) as a function of metallicity. We find that all stars with [Fe/H] $> -1.0$ have [Ba/Y] $< 0$ to within $1\sigma$.  }
    \label{fig:s_process}
\end{figure*}

In Figure \ref{fig:abundances}, we plot log$(g)$ vs. $T_{\rm eff}$ as well as abundances of several $\alpha$ and s-process elements as a function of [Fe\,I/H]. For reference, in the background, we plot distributions of the corresponding parameters for all sources in the GALAH DR4 catalog that satisfy the relevant quality cuts (\texttt{flag\_X\_fe} = 0, where \texttt{X} is the element of interest, as well as \texttt{flag\_sp} = 0 and \texttt{flag\_fe\_h} = 0). Systems with best-fit dark secondary masses below and above $1.4\,M_{\odot}$ are plotted in blue and red, respectively. 

Our sample lies on the expected region for MS stars in panel (A), with more metal-poor stars being hotter on average. 

Panels (B) to (D) show $\alpha$ element abundances relative to iron, [$\alpha$/Fe], against [Fe\,I/H]. Our measured abundances follow the typical anti-correlation between [$\alpha$/Fe] and [Fe/H] observed in the Galactic disk \citep[e.g.][]{Hayden2015ApJ}, with a high-$\alpha$ plateau at low [Fe/H] due to rapid enrichment of stars formed at early times by core collapse supernovae \citep[e.g.][]{Burbidge1957RvMP, Tinsley1979ApJ, Tinsley1980FCPh,  DelgadoMena2017A&A, Buder2021MNRAS}. This is one indication that our measured abundances are reasonable. 

Lastly, in panels (E) and (F), we show abundances of singly ionized barium and yttrium. We see more complex structures in these panels that diverge from the trends seen for typical stars in GALAH DR4. In general, our sample has a larger fraction of stars that are enriched in these s-process elements. This is expected, since our targets were specifically selected to have WD companions and a possible history of accretion from an AGB companion.

In the left panel of Figure \ref{fig:s_process}, we see that as expected, there is a positive correlation between the [Ba/Fe] and [Y/Fe], which strengthens at larger values. The best-fit line obtained using orthogonal distance regression has a slope $(0.731 \pm 0.055)$ and intercept $(0.166 \pm 0.022)\,$dex. While the literature varies in the exact definition of ``barium star/dwarf", most such stars have [$s$/Fe]$\gtrsim 0.2-0.3$ (where $s$ is an s-process element; \citealt[e.g.][]{deCastro2016MNRAS, Escorza2023A&A, Rekhi2024ApJL, Rekhi2025arXiv}). In this work, we classify all stars with [Ba/Fe]$>0.25$ as s-process enhanced (i.e. barium dwarfs). For stars which do not have a measured [Ba/Fe], we consider [Y/Fe]$>0.35$, based on the linear fit in Figure \ref{fig:s_process}).

Under this definition, a total of 43 systems in our sample are classified as s-process enhanced. 4 of these were previously identified by \citet{Rekhi2024ApJL}, who cross-matched the astrometric WD + MS binary catalog to GALAH DR3, while the remaining 39 are new discoveries. The typical uncertainty on [Ba/Fe] and [Y/Fe] is $\sim 0.05\,$dex. If we enforce stricter $1\sigma$ criteria on the minimum abundances (e.g. [Ba/Fe]$>0.25+\sigma$), the total number of new discoveries decreases from 39 to 34. As the exact number depends on the adopted definition, and to remain consistent with \citet{Rekhi2024ApJL}, we take the numbers that do not incorporate errors as our fiducial values.

In the right panel of Figure \ref{fig:s_process}, we plot [Ba/Y] (i.e. [Ba/Fe] - [Y/Fe]) as a function of the metallicity, [Fe/H]. This is a measure of the relative abundance of heavy (e.g. Ba, La, Ce) to light s-process (e.g. Y, Zr) elements (often denoted as [hs/ls]), which is commonly used to compare observations to predicted yields from AGB nucleosynthetic models \citep[e.g.][]{Lugaro2003ApJ, Allen2006A&A}. Above [Fe/H]$=-1.0$, almost all stars are consistent with negative [Ba/Y] within their observational uncertainties, while the two stars with [Fe/H]$<-1.0$ have positive [Ba/Y]. This implies that the yields of the light s-process element, Y, is greater than heavy s-process element, Ba, in AGB donors, unless they are very metal-poor. Although this trend between [Fe/H] and [Ba/Y] is in the expected direction based on previous work, we note that it is weaker than found in most previous work, in particular below [Fe/H]$=-1.0$ \citep[e.g.][]{Smiljanic2007A&A, Cristallo2016JPhCS, Cseh2018A&A, Yang2024MNRAS}. 

\begin{figure*}
    \centering
    \includegraphics[width=0.95\linewidth]{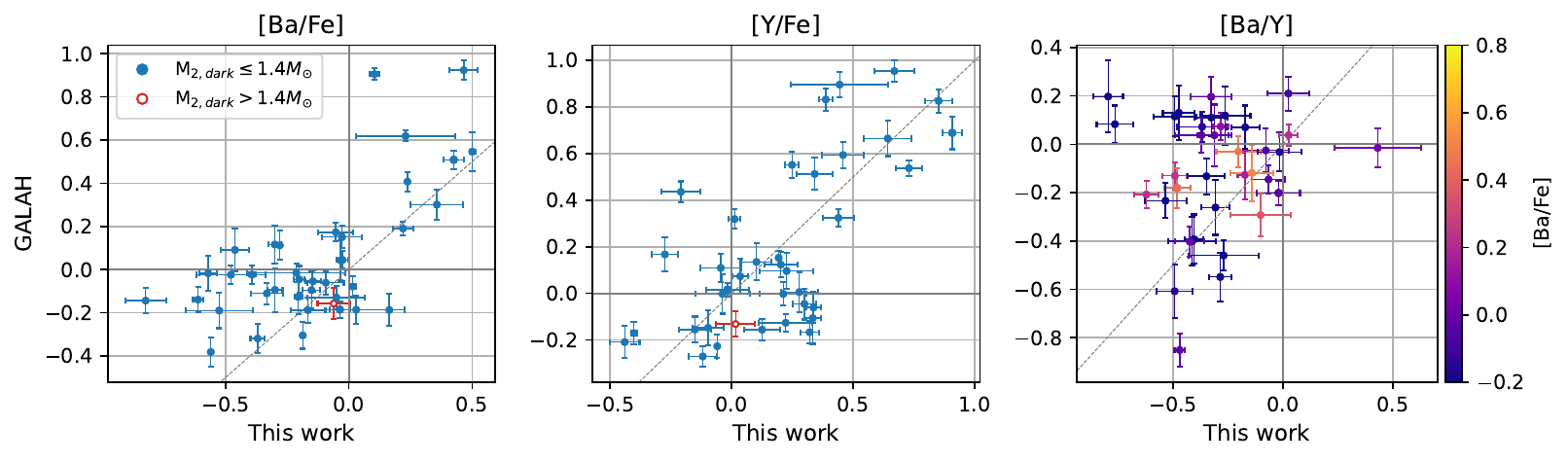}
    \caption{Comparison between our derived Ba and Y abundances, and their differences, against those in the GALAH DR4 catalog. In the rightmost panel, points are colored by [Ba/Fe]. Our abundance measurements are generally correlated with those from GALAH, but many stars are not consistent within their uncertainties. Nevertheless, we agree in almost all cases about whether or not a target is classified as a barium star.}
    \label{fig:abun_galah}
\end{figure*}

In Figure \ref{fig:abun_galah}, we compare our derived s-process abundances to those from GALAH DR4 (where available). There is an expected correlation between measurements of [Ba/Fe] and [Y/Fe], with increasing spread at lower abundances where line strengths are weaker. In the right panel, we compare the [Ba/Y] measurements. There is a subset of systems occupying the upper left corner for which our measured [Ba/Y] are significantly lower than those from GALAH. However, these are not barium dwarfs. Within the sample that overlaps with GALAH, all of the objects we classify as barium dwarfs would also be classified as such based on their GALAH abundances. On the other hand, four objects classified to be barium dwarfs using GALAH are excluded with our abundances. 

We find no clear separation in the abundance patterns for systems in the NCE vs. RCE samples, which suggests that -- within our sample, which is restricted to RCE stars with relatively small color excesses, low eccentricity, and companion masses near $0.6\,M_{\odot}$ -- they probe the same underlying population. In particular, there are barium dwarfs in the RCE sample, which would not be expected if they were triples containing tight MS+MS binaries. This reflects the fact that the cut on color excess removes some fraction of true WD + MS binaries from the non-class I sample, as discussed in \citet{Yamaguchi2025arXiv}.

\subsection{Dependence on orbital and stellar parameters} \label{ssec:param_dependence}

\begin{figure*}
    \centering
    \includegraphics[width=0.95\linewidth]{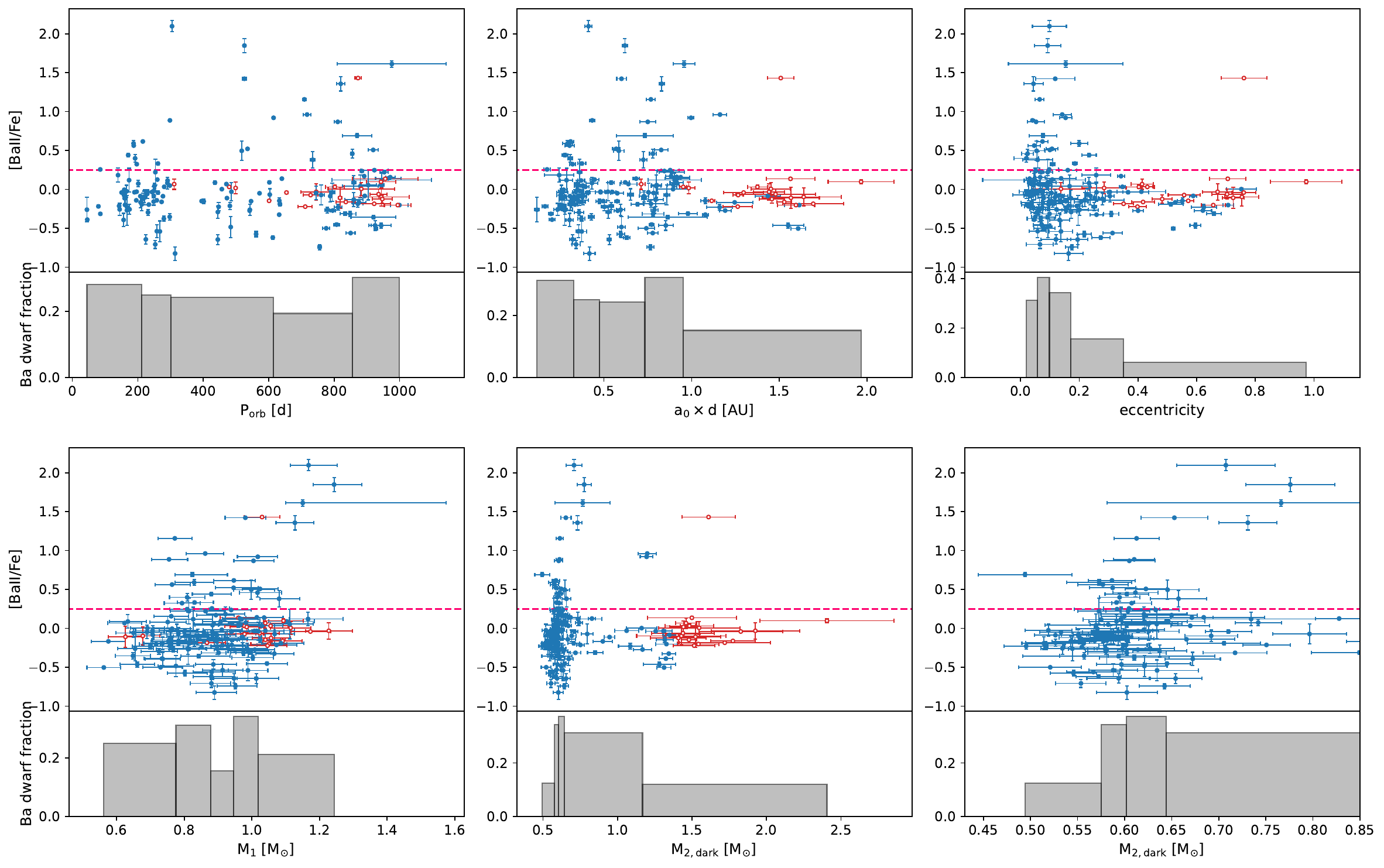}
    \caption{Barium abundance (\textit{top}) and barium star fraction (\textit{bottom}) as a function of several orbital and stellar parameters: (A) orbital period, (B) photocenter semi-major axis, (C) eccentricity, (D) primary mass, and (E.1, E.2) dark companion mass. Panel (E.2) is a zoom-in of panel (E.1) for companion masses below $0.85\,M_{\odot}$. The pink dashed line indicates the threshold for barium dwarfs, [Ba/Fe]$=0.25\,$dex.}
    \label{fig:baII_params}
\end{figure*}

In Figure \ref{fig:baII_params}, we show the measured [Ba/Fe] against several orbital and stellar parameters: orbital period ($P_{\rm orb}$), photocenter semi-major axis ($a_0 \times d$, where $a_0$ is the angular size in mas and $d$ is the distance in kpc; this is equal to the primary semi-major axis for a dark secondary), eccentricity ($e$), MS star mass ($M_{\rm 1}$), and companion mass ($M_{\rm 2, dark}$). $M_{\rm 1}$ was obtained from Gaia's \texttt{binary\_masses} catalog. $M_{\rm 2, dark}$ is the mass inferred from the astrometric orbit assuming a completely dark companion, which is a reasonable assumption for a cool WD or NS. To estimate the uncertainty, we use Monte Carlo sampling, generating 100 realizations of $M_{\rm 2, dark}$ by drawing from Gaussian distributions of the input parameters ($a_0$, $P_{\rm orb}$, $M_{\rm 1}$, and parallax $\varpi$) and taking the standard deviation of these realizations. The histogram under each plot shows the fraction of barium dwarfs (as defined in Section \ref{ssec:results_abundance_patterns}) over the total number of stars in our sample. We selected bins so that the number of systems in each bin is approximately constant. The same plots for [Y/Fe] can be found in Figure \ref{fig:yII_params} in Appendix \ref{appendix:yII_trends}. 

Overall, we see no strong dependence of either [Ba/Fe] or the barium dwarf fraction with orbital period or semi-major axis. There is no strong trend with eccentricity at $e < 0.3$, but at higher eccentricities, we find only one barium star (Section \ref{ssec:ba_star_ns}). This is likely due to highly eccentric systems hosting NS or high-mass WD secondaries, which originate from more massive progenitors with lower s-process yields (Section \ref{sec:theory}). 

In panels (D) and (E.2), we see that for barium dwarfs, [Ba/Fe] increases with both increasing primary and secondary mass, consistent with the findings of \citet{Rekhi2024ApJL, Rekhi2025arXiv}. It is important to point out that these quantities are not independent of each other. More massive primaries require more massive WDs to rule out single MS companions and enter the non-class I sample. 

However, the trend with secondary mass eventually falls off for masses above $\sim 0.8\,M_{\odot}$. Moreover, across the entire range of companion masses, there are stars which show no s-process enhancement. This suggests that mass is not the sole determining factor for whether or not stars end up as barium dwarfs. We discuss possible physical explanations of the observed trends and scatter in Section \ref{sec:theory}. 

\subsection{Ba enhancement in a NS companion?} \label{ssec:ba_star_ns}

One barium dwarf is of particular interest (Gaia DR3 source ID 1386979565629462912). It is metal-poor, with [Fe/H]$\sim -1.6$, and is strongly barium enhanced, with [Ba/Fe]$=1.43\,$dex. The latter is derived from a single Ba\,II line at $5854\,$\AA, which we have confirmed to be well-fitted by the Gaussian model. Unfortunately, its HIRES spectrum does not contain regions of Y\,II lines. It is the only barium dwarf in our sample with secondary mass above the Chandrasekhar limit ($1.6\pm0.2\,M_{\odot}$; Panel (E.1) in Figure \ref{fig:baII_params}). It has an orbital period of $\sim 870\,$d and is in a highly eccentric orbit, with $e = 0.76\pm0.08$. Additionally, it is classified by \citet{Shahaf2024MNRAS} as an NCE source. These properties all point to a NS companion. However, this would imply a progenitor star with a mass above $\sim 8\,M_{\odot}$ which would have evolved into a super-AGB (SAGB) star. The interiors of SAGB stars are thought to have different physical conditions from lower-mass AGB stars, which are expected to affect yields of s-process elements. Theoretical works of SAGB stars generally predict low yields of heavy s-process elements like barium, though model uncertainties associated with mixing processes and nuclear reaction rates can affect precise yields \citep[e.g.][]{Siess2010A&A, Lau2011ASPC, Doherty2017PASA}. Additional follow-up spectroscopy with better wavelength coverage to allow measurements of light s-process elements, as well as tracers of neutron density such as rubidium, will allow us to make more robust comparisons with theoretical expectations. 

Additionally, this object shows evidence of carbon enhancement (Section \ref{ssec:cemp_stars}), which is also in tension with a massive SAGB origin. While dredge-up episodes in AGB stars are expected to bring up carbon to the outer layers and enrich them, this is thought to be avoided in SAGB stars, where the base of their convective envelopes reach sufficiently high temperatures to destroy carbon in a process known as ``hot bottom burning" \citep[e.g.][]{Ventura2013MNRAS, Doherty2014MNRAS, DellAgli2017MNRAS}.  

We caution that the Gaia orbit of this source has not yet been validated \citep[e.g.][]{El-Badry2024OJAp}. It is possible that the implied secondary mass is overestimated, meaning that we cannot rule out the possibility of a massive WD secondary. We note that primary masses reported in the \texttt{binary\_masses} table for metal-poor stars tend be overestimated \citep{El-Badry2024OJAp_NS}. Multi-epoch spectroscopic observations to measure and fit RVs will allow us to strengthen constraints on the orbit and nature of the companion. If the abundances and companion mass are reliable, it is possible that the companion is a merger product of two lower-mass WDs \citep[e.g.][]{Hallakoun2026}. 

One other barium dwarf has a notably massive companion (Gaia DR3 source ID 5648541293198448512). However, it has a relatively low implied barium abundance ([Ba/Fe] = $0.26$) and little to no yttrium enhancement, and the large error on its secondary mass makes it consistent with being below the Chandrasekhar limit to within $1-\sigma$ ($1.5\pm 0.3\,M_{\odot}$). While this makes it a weaker NS candidate with a less robust detection of s-process enhancement, it is in a highly eccentric ($e\sim 0.7$) orbit and may still warrant additional observations. 

\subsection{Carbon enhanced barium dwarfs} \label{ssec:cemp_stars}

\begin{figure*}
    \centering
    \includegraphics[width=0.95\linewidth]{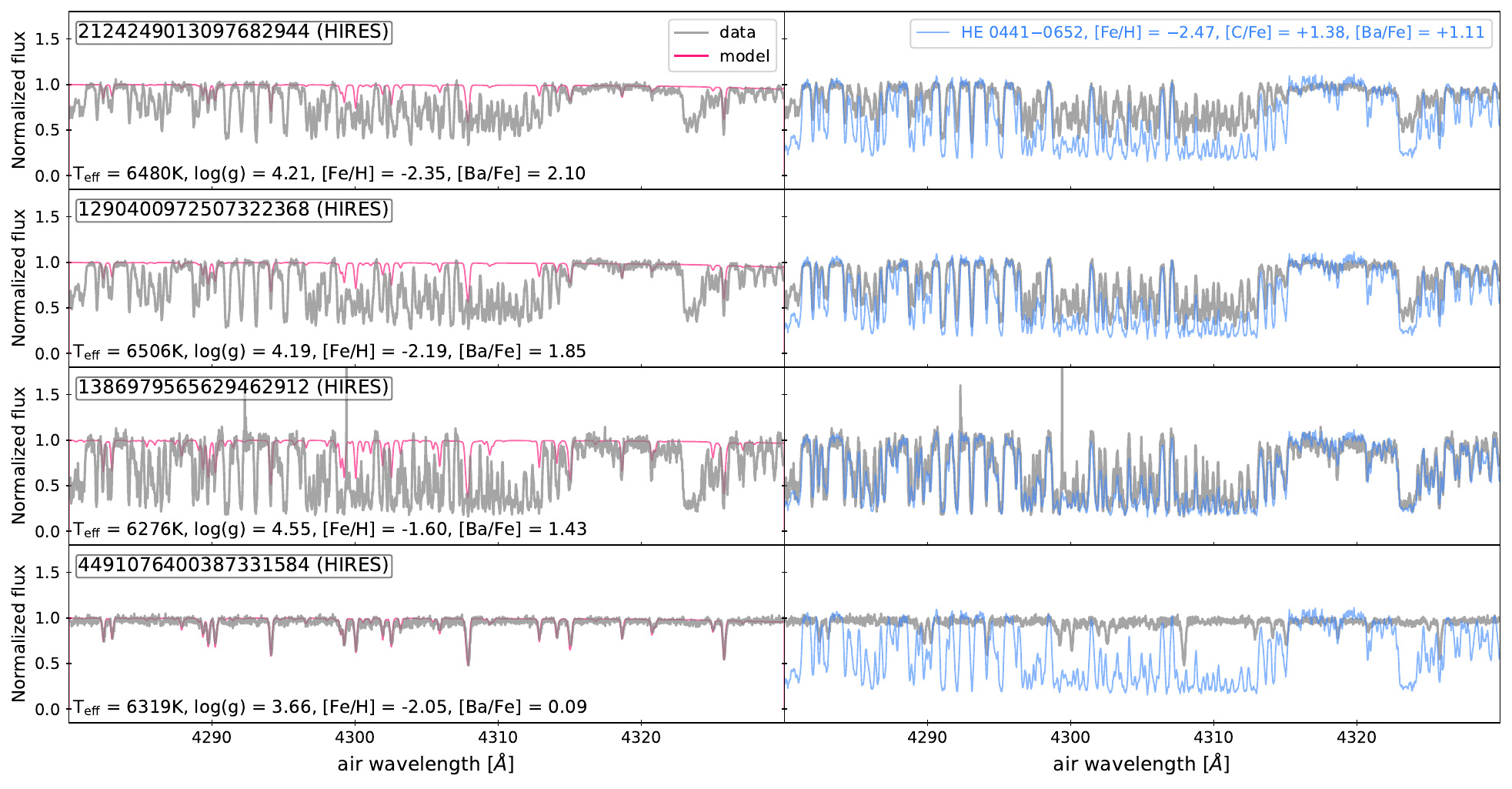}
    \caption{CH G-band region of some of the most metal-poor stars in our sample, all with [Fe/H]$\lesssim-1.6$ (gray). The first three stars are barium dwarfs, while the fourth is not. \textit{Left}: In pink, we plot synthetic spectra generated using the best-fit atmospheric parameters corresponding to each star (Section \ref{ssec:spectral_fitting}). \textit{Right}: In blue is the UVES spectrum of a literature CEMP-s star, HE 0441-0652 \citep{Hansen2016A&A, Zhang2011A&A}. The deep absorption features of the three barium dwarfs are absent in their synthetic spectra and match those of the CEMP-s star, suggesting that they are carbon enhanced. In contract, there is no evidence of such an enhancement in the fourth star.}
    \label{fig:ch_g_band}
\end{figure*}

Three barium dwarfs, which are among the most metal-poor in our sample ([Fe/H]$\lesssim-1.6$), show deep absorption bands located around $\sim 4300\,$\AA \ that are not predicted by their synthetic spectra (Section \ref{ssec:spectral_fitting}). We show this in the left column of Figure \ref{fig:ch_g_band}. For comparison, we plot the same region for an additional star where this absorption is absent and the synthetic spectrum provides a good fit to the data. This star is similarly metal-poor to the other three ([Fe/H]$\sim-2.0$), but is not barium enhanced. 

We identify this absorption band as the G band of CH. This, along with the $\rm C_2$ Swan band (also visible in the spectra of the three stars; see top row of Figure \ref{fig:synspecs}), is among the only strong carbon features for metal-poor stars at these effective temperatures and therefore is commonly used in the study of CEMP stars \citep[e.g.][]{Goswami2006MNRAS, Spite2013A&A, Hansen2016A&A_2, Norris2019ApJ}. In the right column of Figure \ref{fig:ch_g_band}, we over-plot the spectrum of a known CEMP-s star, HE 0441-0652, from Ultraviolet and Visual Echelle Spectrograph (UVES; \citealt{Hansen2016A&A, Zhang2011A&A}; ESO proposal number 170.D-0010). The deep absorption of the three barium dwarfs match closely to those of the CEMP-s star. The fact that these features are absent in the star with no barium enhancement supports the idea that the both carbon and s-process elements originate from the same source -- a WD companion that was formerly an AGB star. While we leave a more detailed analysis of the carbon abundance for future work, this suggests a smooth transition between the ``barium dwarfs" and ``CEMP-s" classifications (Section \ref{sec:literature}).

\subsection{Sensitivity to the assumed stellar parameters and metallicity} \label{ssec:GALAH_param_tests}

In Section \ref{ssec:spectral_fitting}, we compared stellar parameters derived in this work to those from GALAH and found a systematic offset in [Fe/H]  of $\sim 0.2\,$dex (Figure \ref{fig:sp_comparison}). While discrepancies between results from different methods are to be expected, here, we discuss their possible effects on the calculated s-process abundances.

If we naively take [Fe/H] values from GALAH as a reference for our measured [Ba/H], the calculated [Ba/Fe] (and similarly, [Y/Fe]) decreases. In this case, we find that out of the nine stars in our sample ultimately classified to be barium dwarfs that are also found in GALAH, six remain. All nine stars have FEROS spectra, with lower SNR than typical sources in our sample, so extrapolation of this fraction to the full sample is likely to yield a pessimistic estimate of the barium dwarf fraction.

On the other hand, if we fix all stellar parameters ($T_{\rm eff}$, [Fe/H], log$(g)$) to those from GALAH and re-calculate our abundances, we find that the offsets in other parameters lead to a general increase in [Ba/H] that wins out over the higher metallicities. This means that there is a net increase in [Ba/Fe] and the number of stars classified to be barium dwarfs rises from 9 to 16. While both of these tests introduce internal inconsistencies in the method and therefore should not be taken at face value, they illustrate that the offsets here do not necessarily imply an overestimate in the fiducial barium dwarf fraction found in this work.

\section{Comparison to literature Ba dwarfs} \label{sec:literature}

\begin{figure*}
    \centering
    \includegraphics[width=0.95\linewidth]{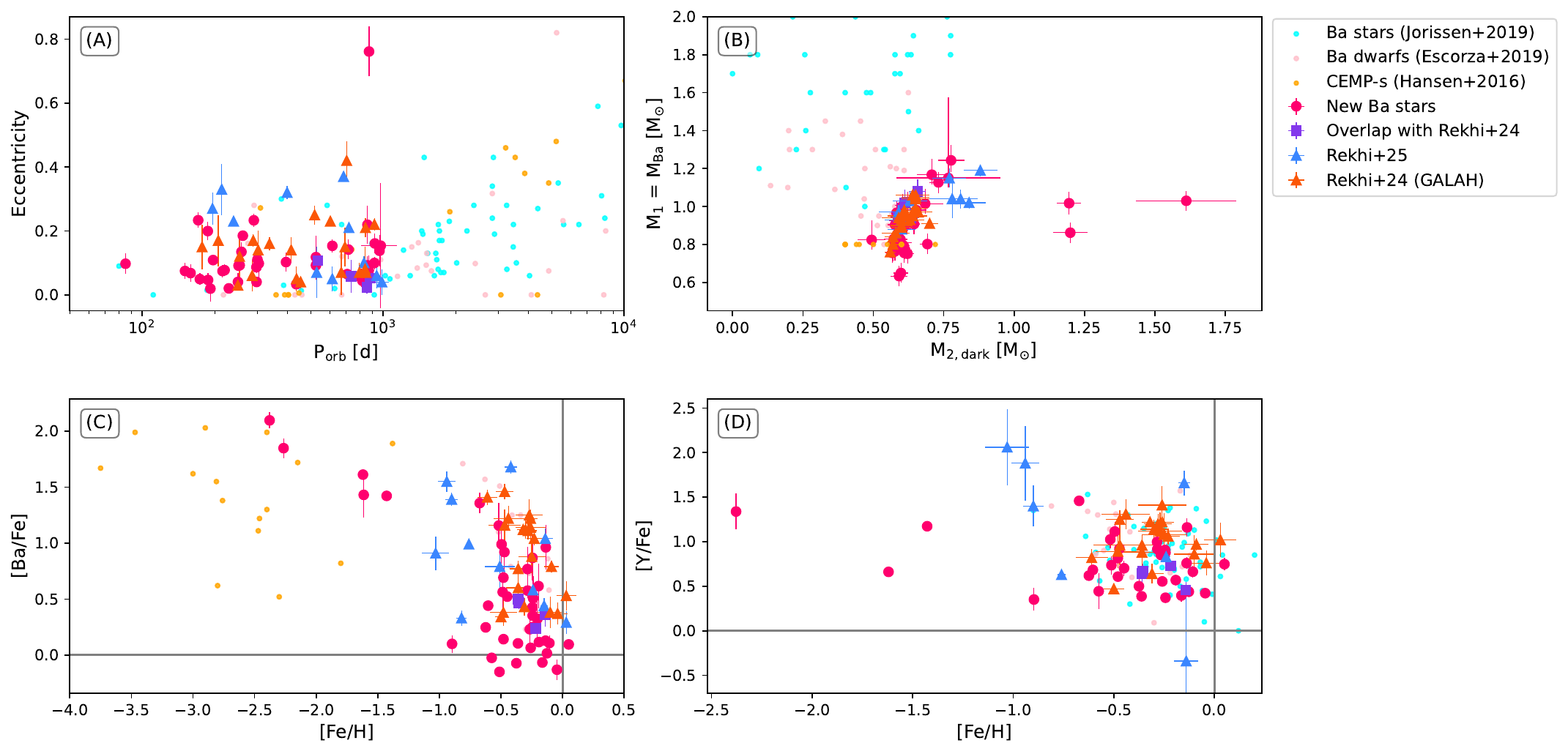}
    \caption{Barium dwarfs from this work and from the published literature, plotted in the space of: (A) eccentricity vs. orbital period, (B) primary vs. secondary mass, (C) [Ba/Fe] vs. [Fe/H], and (D) [Y/Fe] vs. [Fe/H] . We include the astrometric WD + barium dwarf binaries from \citet{Rekhi2025arXiv}, barium stars/giants and dwarfs with spectroscopic orbits from \citet{Jorissen2019A&A} and \citet{Escorza2019A&A} respectively, and CEMP-s stars \citep{Hansen2016A&A}. Note that for the barium stars and giants, the secondary mass plotted is the minimum implied mass for an edge-on inclination. Meanwhile, \citet{Hansen2016A&A} assumes a fixed primary mass of $0.8\,M_{\odot}$ and use this to derive the companion mass which we directly plot here. The systems from this work and \citet{Rekhi2025arXiv} reside in a similar region of these plots, but our additions extend the tails of the distributions. In particular, the most metal-poor systems in our sample bridge the gap between normal barium dwarfs and metal-poor CEMP-s stars.}
    \label{fig:rekhi_comparison}
\end{figure*}

As discussed in \citet{Rekhi2025arXiv}, the Gaia sample of WD + MS binaries contains barium dwarfs in a region of parameter space that was previously sparsely populated. In Figure \ref{fig:rekhi_comparison}, we compare several orbital and stellar properties of our systems containing barium dwarfs to those from \citet{Rekhi2025arXiv}. Where available, we also include barium dwarfs and giants from the literature \citep{Jorissen2019A&A, Escorza2019A&A} as well as CEMP-s stars \citep{Hansen2016A&A}. Since the targets in our work and \citet{Rekhi2025arXiv} are all selected from the same underlying Gaia sample, on average, they occupy a similar region of the parameter space. However, several additional barium dwarfs from this work extend the tails of the distributions, strengthening previously observed trends and introducing interesting outliers. 

In particular, all systems from this work and \citet{Rekhi2025arXiv} have periods ranging from $\sim 100$ to $1000\,$d, which correspond to AU-scale separations, and most have eccentricities between 0 and 0.2. There is a positive relation between the component masses, as the astrometric triage technique requires more massive, and thus luminous, primaries to host more massive WDs \citep{Shahaf2019MNRAS, Shahaf2024MNRAS}. However, as highlighted in Section \ref{ssec:ba_star_ns}, we find two barium dwarfs around massive secondaries very close to or exceeding the Chandrasekhar limit, and these correspond to the two most eccentric systems with $e\sim0.7$, making them consistent with having NS companions. In addition, there are two barium dwarfs around ultra-massive WD candidates with implied masses $\sim 1.2\,M_{\odot}$. It is worth mentioning that these systems have relatively low eccentricities ($\sim 0.2$), consistent with the bulk of the sample and unlike the two NS + MS binary candidates.

There is an inverse trend between the barium abundance and metallicity. This trend is strengthened by the additions from our work, which extends the range of [Fe/H] from the \citet{Rekhi2025arXiv} sample down by more than 1 dex.

Note that while the systems studied this work and \citet{Rekhi2024ApJL} were selected from the same underlying population from \citet{Shahaf2024MNRAS}, \citet{Rekhi2024ApJL} preferentially selected some targets based on having high eccentricities, and thus the different samples have heterogeneous selection functions. In addition, since the other literature barium dwarfs/stars were not selected based on their AMRFs (Section \ref{sec:selection}), they are on average more massive and luminous than those of our sample or \citet{Rekhi2025arXiv}. 

As described in Section \ref{ssec:cemp_stars}, we identified evidence of carbon enhancement in a few of the most metal-poor barium dwarfs in our sample. Indeed, these overlap in the space of [Ba/Fe] vs. [Fe/H] with the literature CEMP-s stars. There is no clear dividing line between the two populations, which suggests that they share a common physical origin. 

If both barium and carbon indeed originate from an AGB companion, it is natural to ask why carbon enhancement is not present in all barium dwarfs, but only the most metal-poor ones. While relative enhancement of any accreted element would be greater at lower metallicity, it is unclear why the strength of this effect varies between elements. One possible reason is the higher oxygen abundances in more metal-rich accretors. CEMP-s stars generally have high carbon to oxygen ratios ($\mathrm{[C/O]}\gtrsim1$). In the presence of oxygen, carbon binds to it to form carbon monoxide (CO) which, unlike CH or C$_2$, does not have strong spectral features in the optical \citep[e.g.][]{Lambert1986ApJS, Beers2007AJ, Kennedy2011AJ, Skuladottir2015A&A, Gonneau2016A&A}. This may explain why CH and C$_2$ absorption features are only found in the most metal-poor stars, where oxygen abundances are low. This, however, depends on CO molecules surviving in the atmospheres of these stars, which will not be true given high enough effective temperatures. Another possibility is that the accreted material is more carbon-rich at lower metallicities. Indeed, predictions of the carbon abundance, as well as the relative abundance of carbon to oxygen, in theoretical models of AGB stars increases with decreasing metallicity. This is shown in Section \ref{sec:theory} (Figure \ref{fig:c_fruity}). 

\section{What determines s-process enhancement?} \label{sec:theory}

\begin{figure*}
    \centering
    \includegraphics[width=0.95\linewidth]{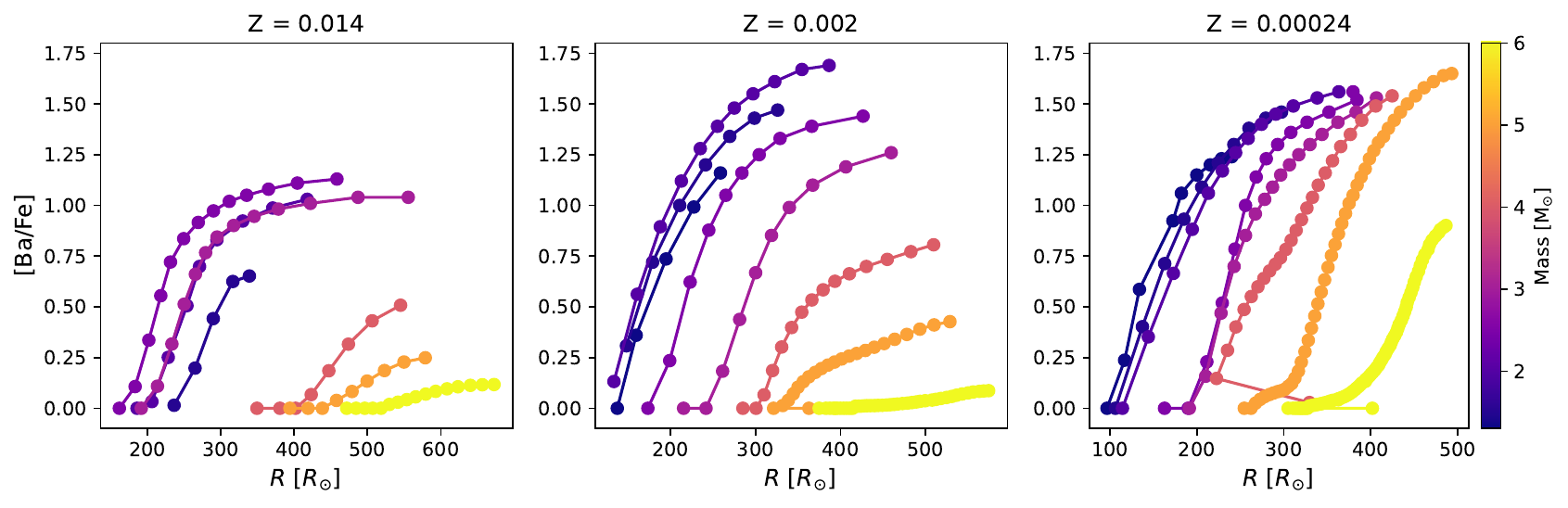}
    \caption{Predicted barium abundance of AGB star envelopes as a function of radius from the FRUITY models. We show models for initial masses ranging from 1.3 to 6.0$\,M_{\odot}$ at three different metallicities. At a fixed radius, the barium yield generally falls with increasing mass and metallicity.}
    \label{fig:fruity}
\end{figure*}

In Section \ref{ssec:param_dependence}, we studied how the s-process abundances of objects in our sample vary with binary parameters. Here, we discuss the physical origins of the trends, and lack thereof, by comparing to theoretical models of the donor AGB stars (i.e. WD progenitors) and the accreting primaries (i.e. MS stars).

In Figure \ref{fig:fruity}, we plot predicted [Ba/Fe] for AGB star envelopes as a function of stellar radius for a range of initial stellar masses and three different metallicties ($Z$). These predictions are from the FRUITY database \citep{Cristallo2009ApJ, Cristallo2015ApJS, Cristallo2016JPhCS}, which provides nucleosynthetic yields and abundances after each dredge-up episode on the AGB, obtained using a modified version of the evolutionary code FRANEC \citep{Chieffi1989ApJS, Chieffi1998ApJ, Straniero2006NuPhA}. A complementary database (ph-FRUITY) provides stellar parameters after every thermal pulse, which can be linked to the dredge-up episodes. We show models with $Z$ = 0.014 (solar), 0.002, and 0.00024, corresponding to [Fe/H] $\sim$ 0, -0.8, and -1.8 which roughly spans the metallicities of our observed systems. 

[Ba/Fe] is predicted to increase as the AGB star evolves and its maximum radius grows, corresponding to more dredge-up episodes and thermal pulses. Thus, the expected enhancement in the accretor due to MT from an AGB donor depends on when MT occurs. Therefore, we may expect a correlation between measured [Ba/Fe] of the MS stars and the binaries' \textit{initial} separations. However, this evidently does not translate to a strong correlation with the present-day separations (Section \ref{ssec:param_dependence}; Figure \ref{fig:baII_params}). This is not unexpected, as MT can change the orbital separation in ways that may depend non-trivially on the binary parameters (see Section \ref{ssec:ba_star_sim} for further discussion). 

The predicted [Ba/Fe] also varies with $Z$. For stars of all masses, [Ba/Fe] generally increases with decreasing $Z$ at a fixed radius. This partially explains the trend between the observed s-process abundances of our barium dwarfs with their metallicities (Section \ref{ssec:results_abundance_patterns}; Figure \ref{fig:abundances}). For masses between $\sim 2.0$ and $4.0\,M_{\odot}$, the trend is broken at early times (i.e. at smaller radii), which may add scatter to the observed trend. 

We can similarly study the evolution of [Ba/Y] as a function of mass and metallicity in the FRUITY models. Overall, this evolution is more complex than that of [Ba/Fe]. It is generally positive for masses below $\sim 4\,M_{\odot}$ and $Z \lesssim 0.014$. For more massive stars, [Ba/Y] evolves non-monotonically, starting at small positive values, then dropping to negative values before increasing again. 

\begin{figure*}
    \centering
    \includegraphics[width=0.95\linewidth]{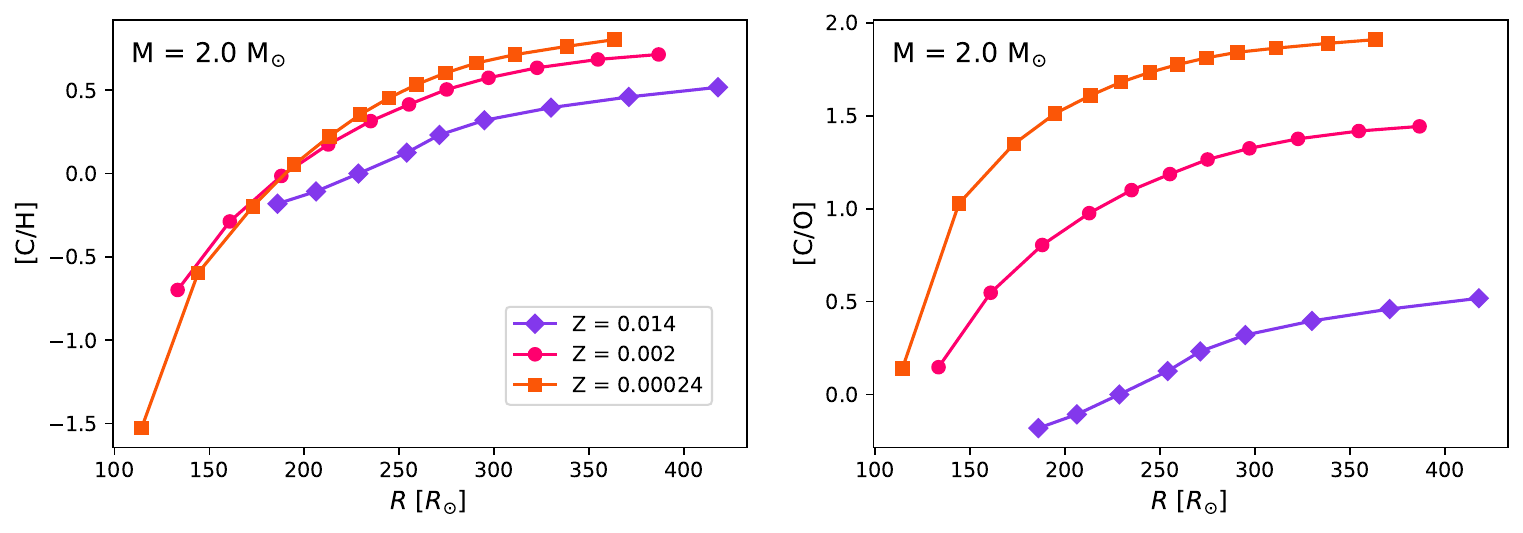}
    \caption{Evolution of [C/H] (\textit{Left}) and [C/O] (\textit{Right}) for an initially $2\,M_{\odot}$ star as predicted by  FRUITY models with a range of metallicities. [C/H], which represents the total carbon yield, increases modestly with decreasing metallicity at fixed radius, but [C/O] increases by almost two dex.}
    \label{fig:c_fruity}
\end{figure*}

Additionally, in the left panel of Figure \ref{fig:c_fruity}, we plot the predicted [C/H] ($\rm=[C/Fe]-[Fe/H]\approx[C/Fe]-log_{10}(Z)$) as a function of stellar radius and metallicity for a fixed initial mass of $2\,M_{\odot}$. [C/H] generally increases with decreasing metallicity, which we may expect to result in more carbon-rich accretors in metal-poor systems. In the right panel, we plot the predicted [C/O] which similarly shows an anti-correlation with metallicity, though more pronounced throughout the AGB evolution. If the carbon is indeed effectively locked up in carbon monoxide, this may be another important explanation for the CEMP stars (Section \ref{sec:literature}). 

\begin{figure}
    \centering
    \includegraphics[width=0.95\linewidth]{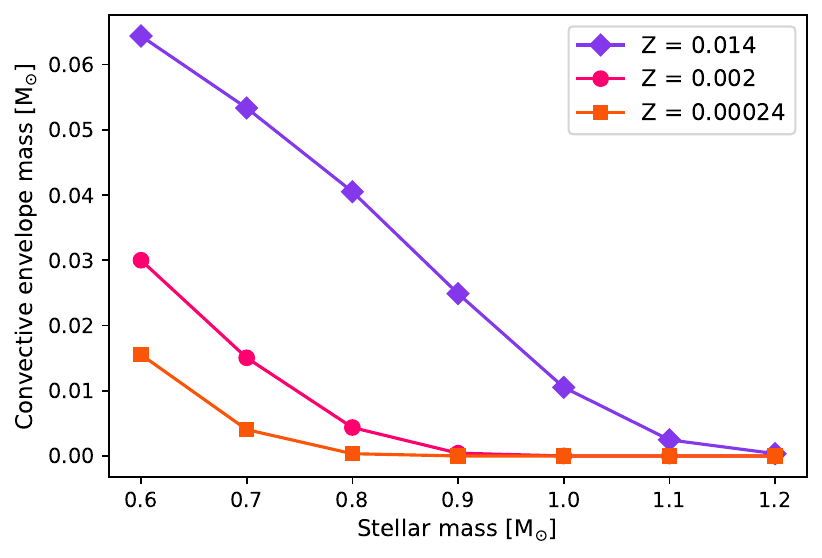}
    \caption{Mass in the convective envelope predicted by MESA models at different metallicities. The convective envelope shrinks with increasing stellar mass and decreasing metallicity. This means that s-process rich material from an AGB donor will experience less dilution, and therefore lead to a stronger surface s-process enhancement, in more massive and metal-poor accretors.}
    \label{fig:mesa}
\end{figure}

Besides the chemical composition of the accreted material, the present-day abundances will also depend on the amount of dilution experienced by this material as a result of mixing in the convective envelope of the accretor. To investigate this, we use the 1D stellar evolution code Modules for Experiments in Stellar Astrophysics (MESA, version r24.03.1; \citealt{Paxton2011, Paxton2013, Paxton2015, Paxton2018, Paxton2019, Jermyn2023ApJS}) to produce MS star models and study changes in the masses of their convective envelopes as a function of the total stellar mass and metallicity. We take the model closest to when the core mass fraction of H reaches $30\%$. Our grid covers a mass range from 0.6 to 1.2$\,M_{\odot}$ to span the range of MS star masses in our sample, and we test the same metallicities as the chosen FRUITY models. We plot the results in Figure \ref{fig:mesa}. We see that that the mass in the convective envelope decreases steeply with both stellar mass and $Z$. At solar metallicity, the mass in the convective envelope becomes negligible above $\sim 1.2\,M_{\odot}$, while at $Z = 0.002$ and $0.00024$, this occurs by $\sim 0.9\,M_{\odot}$. This means that equal amounts of accreted material with the same s-process content will likely lead to stronger enhancement in a more massive and more metal-poor star.

\subsection{Population modeling of Ba dwarf binaries} \label{ssec:ba_star_sim}

Here, we take the theoretical models described above to predict barium abundances in the population of astrometric WD + MS binaries to test whether or not the aforementioned factors are able to account for the observed spread. 

We build on the results of \citet{Yamaguchi2025arXiv}, who used a forward modeling approach to create a mock non-class I sample with properties that closely approximate the true \citet{Shahaf2024MNRAS} sample. They start with a simulated population of Galactic zero-age binaries, then model the stellar evolution of each component as well as their orbital evolution due to binary interactions to obtain the present-day population of WD + MS binaries. In their fiducial model, the change in orbital separation due to MT from AGB donors for systems with initial accretor-to-donor mass ratios exceeding $\sim 0.4$ -- which ultimately become the AU-scale WD + MS binaries that we study in this work -- is calculated using the relations derived by \citet{Soberman1997AA} for fully non-conservative MT (i.e. $\beta = 1$). Epoch astrometry is predicted for each present-day binary based on the Gaia scanning law which is then fit with astrometric models. A subset of these orbital solutions form the catalog of astrometric binaries which reproduces the key features of the true catalog published in Gaia DR3, suggesting that the selection function derived by this method is reasonable \citep{El-Badry2024OJAp_gaiamock}. Finally, as done by \citet{Shahaf2024MNRAS}, non-class I systems are selected based on the AMRF. We refer readers to \citet{Yamaguchi2025arXiv} for an in-depth description of this process. 

For each WD + MS binary in the mock non-class I sample, we know the initial (i.e. pre-MT) stellar and orbital parameters. From the metallicity of the system, mass of the AGB donor (i.e. the WD progenitor), and initial Roche lobe radius, we can interpolate the FRUITY models (Figure \ref{fig:fruity}) to predict the barium abundance in the donor at the onset of MT, $[\mathrm{Ba/Fe}]_{\mathrm{AGB}}$ which is accreted onto the companion. We estimate the mass of the companion's convective envelope, $M_{\rm conv}$, by interpolating our MESA models (Figure \ref{fig:mesa}) at the companion mass and metallicity. 

Assuming efficient mixing, the final abundance in the accretor's convective envelope, $[\mathrm{Ba/Fe}]_f$, is given by: 
\begin{equation}
\log_{10}\left(\frac{10^{[\mathrm{Ba/Fe}]_i} M_{\rm conv} + 10^{[\mathrm{Ba/Fe}]_{\mathrm{AGB}}} M_{\rm acc}}{M_{\rm conv} + M_{\rm acc}} \right)
\end{equation}
where $[\mathrm{Ba/Fe}]_i$ is the initial barium abundance in the MS companion before MT and $M_{\rm acc}$ is the mass of material accreted. This assumes that both components have the same iron abundance, which is reasonable. Since the accretors in our systems have not left the MS, we assume no initial barium enhancement, $[\mathrm{Ba/Fe}]_i = 0$. By default, $M_{\rm acc}$ is taken to be 1\% of the AGB donor's envelope (i.e. donor mass - WD mass) which corresponds, on average, to $\sim 0.015\,M_{\odot}$, but we experiment with the effects of varying the efficiency of mass transfer. Note that while this calculation is inconsistent with the fully non-conservative assumption made in generating the simulated population, the resulting discrepancy is minimal for such small amounts of accreted material. 

\begin{figure*}
    \centering
    \includegraphics[width=0.95\linewidth]{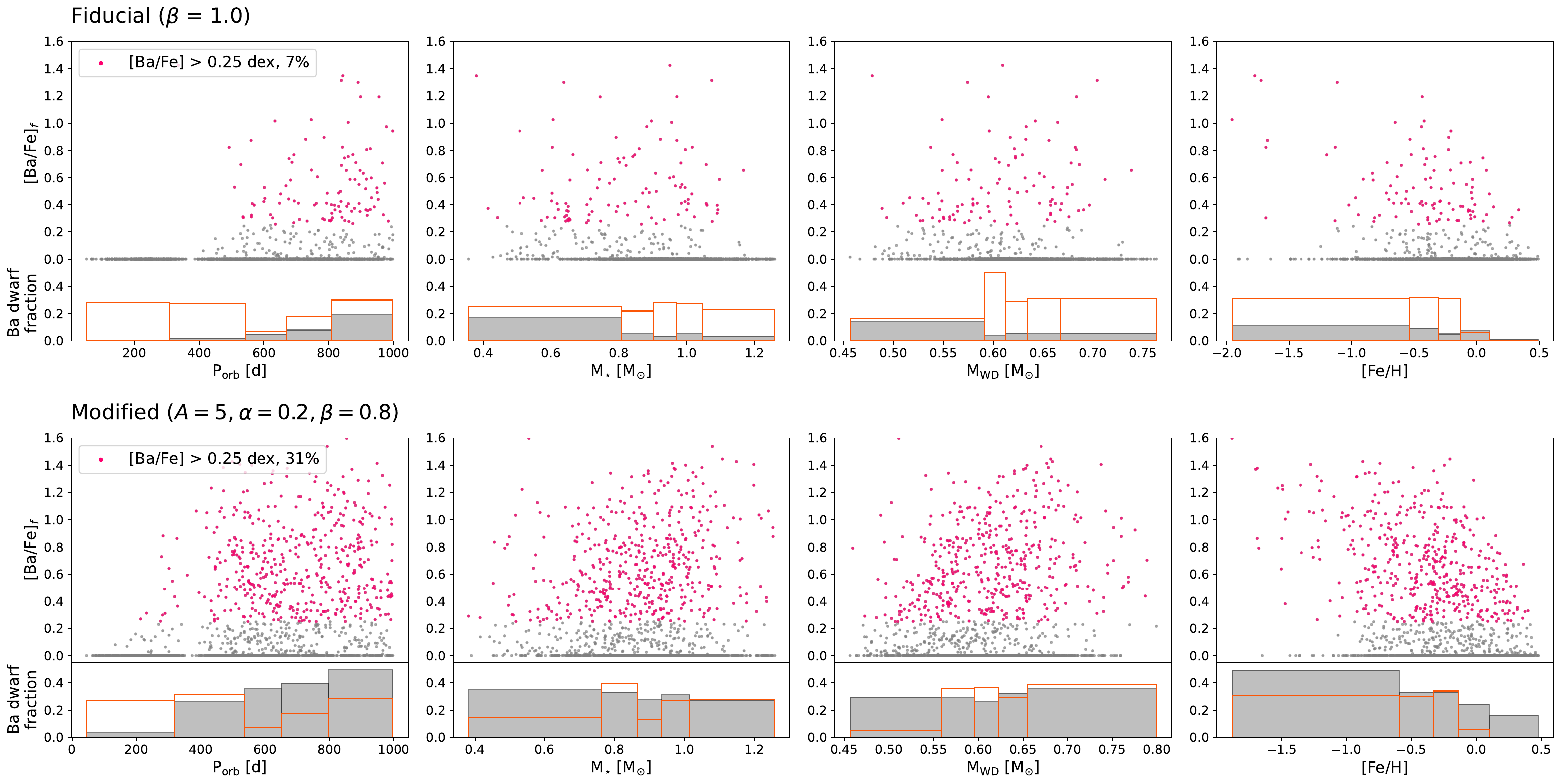}
    \caption{Predicted barium abundance and barium dwarf fraction as a function of several orbital and stellar parameters for the simulated non-class I sample. For visual clarity, we plot a subset of 1500 systems. In magenta, we highlight the barium dwarfs, with [Ba/Fe] $>0.25\,$dex. In the top row are results using the fiducial model of orbital evolution via MT adopted by \citet{Yamaguchi2025arXiv}, using relations from \citet{Soberman1997AA} with $\beta = 1$ (i.e. fully non-conservative MT with mass loss via isotropic re-emission from the accretor). In the bottom row are results  using the same relations, but with non-zero $\alpha$ (i.e. still fully non-conservative, but with some mass loss via winds from the donor). For reference, we over-plot the observed barium dwarf fractions in orange (Figure \ref{fig:baII_params}). On average, the latter model results in more orbital shrinkage, so that systems with initially wider separations and correspondingly more evolved AGB donors enter our sample. As s-process yields increase with time on the AGB, this leads to a larger fraction of barium dwarfs. }
    \label{fig:sim_results}
\end{figure*}

In the top of row of Figure \ref{fig:sim_results}, we plot the predicted $[\mathrm{Ba/Fe}]_f$ as a function of several orbital and stellar parameters for the simulated sample using the fiducial model of orbital evolution adopted by \citet{Yamaguchi2025arXiv}. In total, fewer than 10\% of the simulated MS stars have $[\mathrm{Ba/Fe}]_i >0.25\,$dex. This fraction is primarily determined by the assumed initial separations of the binaries. As shown in Figure \ref{fig:fruity}, [Ba/Fe] increases rapidly with the radius/age of the AGB donor. In the fiducial model, systems with initial Roche lobe radii $\lesssim 200\,R_{\odot}$ comprise the bulk of those that end up as AU-scale WD + MS binaries with $P_{\rm orb} <1000\,$d and thus enter our sample. AGB stars at these radii are in very early phases of s-process nucleosynthesis, and thus the accreted material is not yet strongly enriched in barium. 

Conversely, if we adopt a different model such that systems with initially wider orbits experience more orbital shrinkage to form the AU-scale systems, MT would have begun when the AGB donors were more evolved, resulting in a larger fraction of barium dwarfs. As one example, if we continue to assume fully non-conservative MT but with some mass loss via wind from the donor ($A = 5, \alpha=0.2, \beta=0.8$; \citealt{Soberman1997AA}), our systems would have experienced MT from AGB donors with radii as large as $\sim 400\,R_{\odot}$. This increases the predicted fraction of barium dwarfs to $\sim 30\,$\%. We show the results of this model in the bottom row of Figure \ref{fig:sim_results}. Although the relation between initial to final separation is not one-to-one, there is a general trend of $[\mathrm{Ba/Fe}]_f$ with the orbital period in the simulation, with a smaller fraction of barium dwarfs below $\sim 1\,$yr. We do not find evidence of such a trend in our observed systems (over-plotted with orange empty bars; also see Figure \ref{fig:baII_params}). This likely reflects the fact that the true orbital evolution is more diverse and complex than assumed by our model, meaning there is more scatter in the relation between initial and final separations. Additionally, most classical barium stars/dwarfs are at periods $>1000\,$d, so the fraction may still increase at longer periods. 

In the bottom row, $[\mathrm{Ba/Fe}]_f$ generally rises with increasing component masses and decreasing metallicity. This is primarily the effect of dilution, combined with the correlation between the component masses for systems that enter the non-class I sample and the dependence of predicted yields on these parameters. We see qualitatively similar trends in our observed systems (Figure \ref{fig:baII_params}). However, the same trends are absent or weaker for the results plotted in the top row. One reason is that more massive accretors are on average accompanied by more massive AGB donors, and in the fiducial model, the vast majority of donors with masses $\gtrsim 2\,M_{\odot}$ (around MS companions with masses $\gtrsim 0.8\,M_{\odot}$) interact prior to significant s-process element production ($\lesssim 200\,R_{\odot}$). This counteracts the effect of dilution. 

The effect of dilution is strengthened as the mass of material accreted becomes comparable to the mass of the accretor's convective envelope. Therefore, if we instead assume that only 0.5\% of the mass lost by the AGB donor is accreted by the companion (corresponding to $\sim 0.008\,M_{\odot}$), the trends of  $[\mathrm{Ba/Fe}]_f$ with component masses and metallicity -- which are driven by dilution -- become stronger. As expected, the total fraction of barium dwarfs decreases. Conversely, if the fraction accreted is increased to $10\%$, these trends weaken as $M_{\rm acc}\sim0.15\,M_{\odot} \gg M_{\rm conv}$. There is a modest increase in the fraction of barium dwarfs by 5-10\% from the default assumption of $1\%$, depending on the model. However, the magnitude of barium enhancement rises more significantly, with more than double the number of stars having [Ba/Fe]$> 1.0$. 

We do not attempt to further refine our simulation to reproduce the exact features of the observed systems and their measured s-process abundances. For this, we must consider the selection of our targets from the non-class I systems which is not homogeneous across the multiple observing programs. Moreover, there exists degeneracies in the orbital evolution predicted by the \citet{Soberman1997AA} relation for different choices of parameters which are difficult to disentangle. In addition, for simplicity, we only considered dilution in the context of the outermost convective envelope. However, in reality, non-convective mixing processes such as thermohaline mixing and atomic diffusion \citep[e.g.][]{Sevilla2022MNRAS, Behmard2023MNRAS} may have non-negligible effects, particularly as the convective layer becomes very thin at larger stellar masses and lower metallicities. It is also worth mentioning that the $\mathrm{[Ba/Fe]}_i = 0$ assumption is a simplification as it has been shown to depend on [Fe/H] \citep[e.g.][]{Nissen2011A&A, Reggiani2017A&A, Amarsi2020A&A}. However, we have confirmed that such a variation only reduces the total fraction of barium dwarfs in our simulated population by a few percent\footnote{We tested a linear fit to data from \citet{Reggiani2017A&A} and \citet{Nissen2011A&A}: $\mathrm{[Ba/Fe]} = -0.08625\mathrm{[Fe/H]}-0.14492$.}. 

This exercise suggests that a key factor in determining whether or not a star that has accreted material from an AGB donor becomes a barium dwarf is how evolved the donor was when interaction began. This is in turn determined by the initial separation of the binary. Dilution plays a minor role for systems in our sample, most of which contain  MS stars with  thin convective envelopes. As a result, the observed barium abundances do not place strong constraints on the MT efficiency. Overall, in the framework of this model, it is not surprising that many AU-scale WD + MS binaries are \textit{not} enhanced in barium, simply because they interacted relatively early on the AGB. This may at least partially explain the fact that most previously discovered barium stars and related systems are in  wider orbits than those of our sample (Figure \ref{fig:rekhi_comparison}), though we caution that most previously known barium stars are giants, which have deeper convective envelopes than our targets and thus have experienced more dilution of accreted material. Future Gaia data releases will be sensitive to  binaries with longer orbital periods, further testing how the barium star fraction varies with period and enabling a more direct comparison to literature barium star samples.

\section{Conclusions} \label{sec:conclusion}

We obtained high-resolution optical spectroscopy for 178 WD + MS binary candidates selected from the Gaia DR3 catalog of astrometric binaries. These systems have orbital periods of $\sim 100-1000\,$d, corresponding to AU-scale separations, and are thought to be products of MT from AGB donors. However, the exact nature of the MT process remains uncertain.
Through spectral analysis, we constrained stellar parameters of the MS companions and derived chemical abundances, focusing on the s-process elements barium and yttrium. We summarize our key findings below: 
\begin{itemize}
    \item We selected targets from the \citet{Shahaf2024MNRAS} sample of systems classified as likely WD + MS binaries based on their Gaia astrometric orbits. We specifically selected several metal-poor targets with [Fe/H] $< -1.0$ (Section \ref{sec:selection}, Figure \ref{fig:selection}).
    \item We use ionization/excitation equilibrium of iron lines to determine T$_{\rm eff}$, log$(g)$, [Fe/H], and $\xi$. We find general agreement with existing literature values (Section \ref{ssec:sp_ew_balance}, Figure \ref{fig:sp_comparison}). 
    \item We measure abundances of iron (Fe\,I, Fe\,II), $\alpha$-elements (Mg\,I, Ca\,I, Si\,I), and s-process elements (Ba\,II, Y\,II) (Section \ref{ssec:abundances_derivation}, Figure \ref{fig:abundances}). We identify 43 barium dwarfs ([Ba/Fe]$>0.25\,$dex OR [Y/Fe]$>0.35\,$dex), 39 of which are new discoveries (Section \ref{ssec:results_abundance_patterns}).
    \item About 25\% of the likely WD+MS binaries in our sample are enriched in s-process elements. Their s-process abundances are uncorrelated with orbital period, in agreement with previous work. The incidence of s-process enhancement drops at high eccentricities and companion masses above $1.4\,M_{\odot}$, likely reflecting the presence of NS rather than WD companions. At lower companion masses, we find weak positive correlations with the component masses and an inverse correlation with the iron abundance (Section \ref{ssec:param_dependence}, Figure \ref{fig:baII_params}).
    \item We identify two barium dwarfs around companions whose masses may exceed the Chandrasekhar limit and which are in highly eccentric ($e > 0.7$) orbits, making them possible MS + NS binaries. Further high-resolution spectroscopy with increased wavelength coverage containing additional s-process lines is necessary to strengthen constraints on their abundances, and multi-epoch radial velocities are needed to validate their orbits and confirm their nature (Section \ref{ssec:ba_star_ns}). 
    \item We find evidence of strong C enhancement in three out of four of the most metal-poor barium dwarfs with $\mathrm{[Fe/H]}< -1.5$ (Figure \ref{fig:ch_g_band}), suggesting that there exists a smooth transition between barium dwarfs and CEMP-s stars (Section \ref{ssec:cemp_stars}). 
    \item Our sample roughly doubles the current number of barium dwarfs in AU-scale orbits. It extends the observed trend of [Ba/Fe] with [Fe/H] by \citet{Rekhi2025arXiv} to significantly more metal-poor stars with [Fe/H]$< -2.0\,$dex. (Section \ref{sec:literature}, Figure \ref{fig:rekhi_comparison})
    \item Using theoretical AGB nucleosynthetic yields from the FRUITY database and 1D models of the companion MS stars to predict barium abundances in a simulated population of WD + MS binaries, we show that the observed diversity of s-process abundances can be attributed to the range in metallicities and component masses (which affect both the AGB yields and mass in the accretor's convective envelope) and the age of the AGB donor at interaction. We also find that the fraction of stars that are barium dwarfs is sensitive to the assumed orbital evolution via MT. In general, systems for which MT began at initially wide separations with evolved AGB donors end up with greater barium enhancements (Section \ref{sec:theory}, Figure \ref{fig:sim_results}).
\end{itemize}
Further analysis to determine precise C abundances as well as additional spectra covering a wider range of wavelengths and elemental lines would be beneficial follow-up work. Moreover, future data releases from the Gaia mission will build on this sample, extending to longer orbital periods and allowing for a more homogeneous selection of these systems to conduct more complete and detailed population-level studies. 


\begin{acknowledgments}
We thank the anonymous referee for detailed feedback which improved the manuscript. We thank Ana Escorza, Sophie van Eck, Alain Jorissen,  Param Rekhi, Ben Roulston, and Jay Farihi for useful conversations. We are grateful to Yuri Beletsky, Sam Kim, Angela Hempel, and Maren Hempel for observing help.  This research benefited from discussions in the ZTF Theory Network, funded in part by the Gordon and Betty Moore Foundation through Grant GBMF5076.
NY acknowledges support from the Ezoe Memorial Recruit Foundation scholarship. This research was supported by NSF grant AST-2307232.

\end{acknowledgments}

\begin{contribution}

NY was responsible for completing the data analysis and population modeling of the systems in this work as well as writing and submitting the manuscript. KE advised NY throughout the project, from initial conceptualization to editing the manuscript. HR provided expertise on the spectral analysis and ensured that the stellar parameters and abundances obtained were reliable. RA re-analyzed Gaia XP spectra using corrected parallaxes appropriate for our systems, which we used in our analysis. SS created the original sample of astrometric binaries from which our systems were selected. 


\end{contribution}

%
\facilities{Max Planck:2.2m (FEROS), Magellan:Clay (MIKE),  Keck:I (HIRES)}

\software{REvIEW (\url{https://github.com/madeleine-mckenzie/REvIEW}), q2 (\url{https://github.com/astroChasqui/q2}), iSpec (\url{https://www.blancocuaresma.com/s/iSpec/manual/introduction}), gaiamock (\url{https://github.com/kareemelbadry/gaiamock})}


\appendix

\section{Appendix information}

\subsection{Radial velocities} \label{appendix:rv_tables}

We summarize RVs measured for each of our spectra in Table \ref{tab:rvs}. Errors are calculated by \texttt{iSpec} following \citet{Zucker2003}. The full table can be found in the supplementary materials. 

We caution that we do not attempt to account for zero-point offsets between different instruments, which are expected to be of order $1\,{\rm km\,s^{-1}}$. Precision is also limited by the fact that we used a fixed template at solar metallicity for the cross-correlation across all spectra.

\begin{table}[]
    \centering
    \begin{tabular}{c c c c c c}
        \hline
        Gaia DR3 source ID & JD & RV & Instrument \\
        \hline
        \hline
        41408333753757056 & 2460161.0915 & -10.6 $\pm$ 2.1  & HIRESr \\
        220012968211559296 & 2460161.0787 & -24.5 $\pm$ 2.3  & HIRESr \\
        352523017213180544 & 2460879.0566 & 20.0 $\pm$ 2.8  & HIRESb \\
        369419178036177792 & 2460879.1283 & -65.5 $\pm$ 2.5  & HIRESb \\
        383389126103078400 & 2460847.0807 & -148.4 $\pm$ 3.7  & HIRESb \\
         & ... & & \\
        
    \end{tabular}
    \caption{Measured RVs for all spectra. The complete table can be downloaded online.}
    \label{tab:rvs}
\end{table}

\subsection{Stellar parameters from Gaia XP spectra} \label{appendix:xgboost}

In Table \ref{tab:xgboost}, we provide stellar parameters re-calculated using methods from \citet{Andrae2023ApJS} but with parallaxes corrected for binary motion. We used these values as initial guesses in our analysis (Section \ref{ssec:sp_ew_balance}).

\begin{table}[]
    \centering
    \begin{tabular}{c c c c c c}
        \hline
        Gaia DR3 source ID & $T_{\rm eff}$ & log($g$) & [Fe/H] \\
        \hline
        \hline
        2808994137268673152 & 6255 $\pm$ 80 & 4.29 $\pm$ 0.03 & -0.15 $\pm$ 0.08 \\
        2124249013097682944 & 6211 $\pm$ 97 & 4.26 $\pm$ 0.04 & -1.88 $\pm$ 0.16 \\
        2097059980331156480 & 5594 $\pm$ 54 & 4.50 $\pm$ 0.01 & -0.34 $\pm$ 0.05 \\
        2795884385252655232 & 6027 $\pm$ 90 & 4.33 $\pm$ 0.03 & -0.63 $\pm$ 0.08 \\
        2105234402606388352 & 5400 $\pm$ 42 & 4.53 $\pm$ 0.01 & -0.47 $\pm$ 0.06 \\
         & ... & & \\
    \end{tabular}
    \caption{Stellar parameters derived from Gaia XP spectra re-calculated using methods from \citet{Andrae2023ApJS} with parallaxes from the \texttt{nss\_two\_body\_orbit} table. The complete table can be downloaded online.}
    \label{tab:xgboost}
\end{table}

\subsection{Results of stellar parameter determination} \label{appendix:q2_plots}

\begin{figure*}
    \centering
    \includegraphics[width=0.95\linewidth]{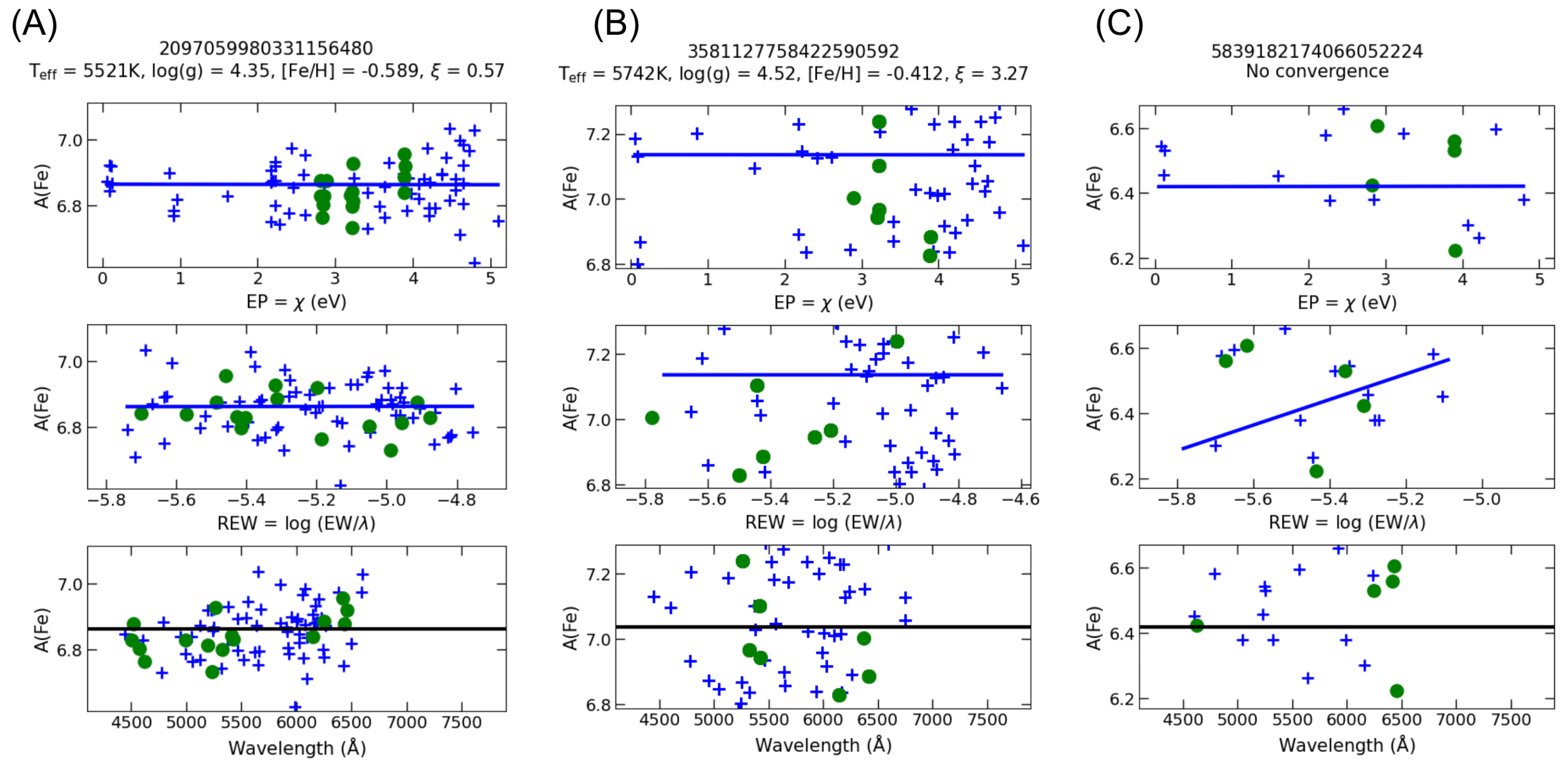}
    \caption{Results of attempting to achieve excitation/ionization equilibrium using iron lines with \texttt{q2} (Section \ref{ssec:sp_ew_balance}). The blue cross markers correspond to Fe I,  while the green circle markers correspond to Fe II lines. The best fit linear model is also plotted in each panel. We show an example of a spectrum which (A) achieves convergence with reliable results on the first attempt, (B) requires fixing log$(g)$ to its initial value to avoid it from going to $5.0$, and (C) fails to achieve convergence due to rotationally broadened lines. Note that the scatter in the data depends on the SNR and resolution of each spectrum.}
    \label{fig:q2_results}
\end{figure*}

In Figure \ref{fig:q2_results}, we show examples of the final output from \texttt{q2} in measuring stellar parameters via excitation/ionization equilibrium for three targets with varying degrees of success (Section \ref{ssec:sp_ew_balance}). Successful convergence is achieved when there are minimal trends in the implied abundances of both Fe\,I and Fe\,II as a function of several line properties, plotted in each of the three rows. Having a small number of lines with reliably measured EWs as a result of poor SNR and/or rotational broadening can lead to convergence failure. For a typical spectrum of our sample, we find that our method is effective for $v\sin(i)\lesssim20\,$km/s. Detailed descriptions for the three examples can be found in the figure caption. 

\subsection{Effect of systematic uncertainty on $T_{\rm eff}$} \label{appendix:teff_err}

\begin{figure*}
    \centering
    \includegraphics[width=0.8\linewidth]{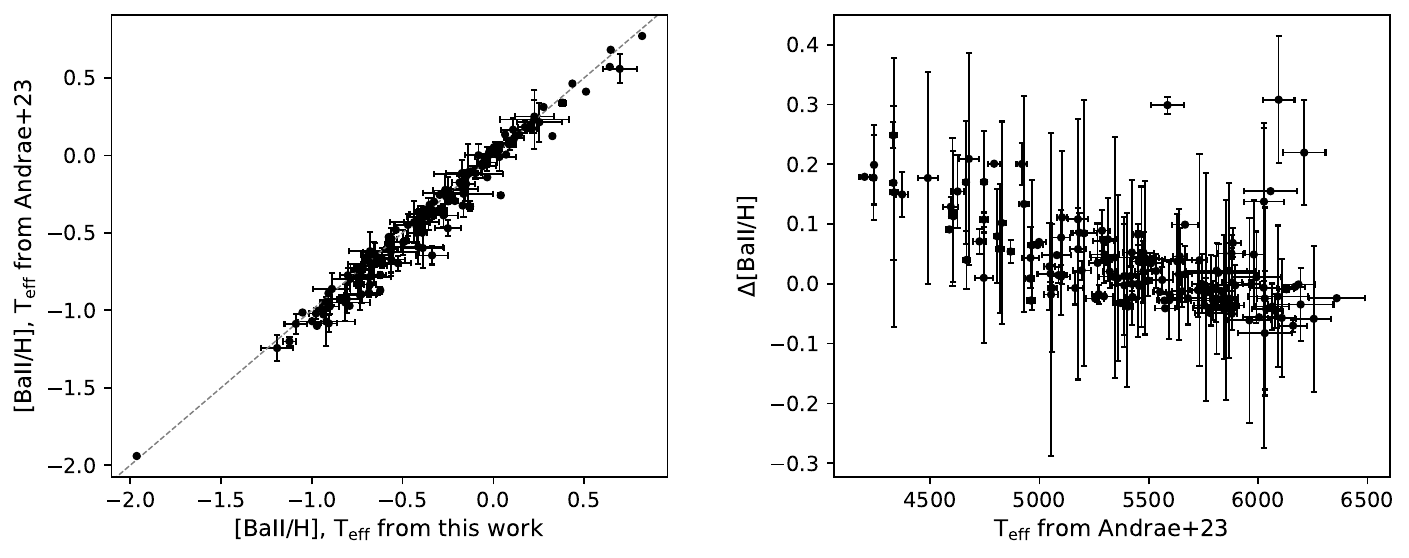}
    \caption{\textit{Left}: Comparison of barium abundances derived taking the best-fit $T_{\rm eff}$ from this work vs. the value from \citet{Andrae2023ApJS} (modified to use  parallaxes corrected for binary motion; Section \ref{ssec:sp_ew_balance}). We take our best-fit values for all other stellar parameters ([Fe/H], log$(g)$, and $\xi$). \textit{Right}: The difference in the abundances as a function of $T_{\rm eff}$. The difference increases for cooler stars, where the two $T_{\rm eff}$ values become increasingly discrepant.}
    \label{fig:teff_test}
\end{figure*}

As discussed in Section \ref{ssec:sp_ew_balance}, the best-fit $T_{\rm eff}$ values derived from high-resolution spectra deviate from those of \citet{Andrae2023ApJS} which are measured using Gaia XP spectra and broad-band photometry. In most cases, the SEDs predicted using $T_{\rm eff}$ values from the latter are more consistent with photometric observations. Nevertheless, we use the spectroscopic stellar parameters to derive our fiducial abundances. Here, we re-derive them using $T_{\rm eff}$ values from \citet{Andrae2023ApJS}, and in Figure \ref{fig:teff_test}, we compare the resulting [Ba/H] (as opposed to [Ba/Fe] which also depends on the iron abundance). We find that our measurements are generally higher, and that this discrepancy increases for cooler stars. However, the discrepancy is less than $0.3\,$dex for the bulk of the sample. Using $T_{\rm eff}$ values from \citet{Andrae2023ApJS} would only reduce the total number of stars classified to be barium dwarfs by one (Gaia DR3 source ID 3474329544020730496, with a [Ba/Fe] close to the lower bound of 0.25\,dex). 

\subsection{Tables of atmospheric parameters and abundances} \label{appendix:full_tables}

In Tables \ref{tab:stellar_params} and \ref{tab:abuns}, we provide snippets of tables summarizing the derived atmospheric parameters and abundances. The full tables containing information for all 160 stars in our final sample can be downloaded from the online journal. 

\begin{table}[]
    \centering
    \begin{tabular}{c c c c c c}
        \hline
        Gaia DR3 source ID & Instrument & $T_{\rm eff}$ [K] & log$(g)$ & $\xi$ [kms$^{-1}$] & [Fe/H] \\
        \hline
        \hline
        2808994137268673152 & HIRESb & 6064 $\pm$ 307 & 4.80 $\pm$ 0.76 & 2.71 $\pm$ 1.61 & -0.29 $\pm$ 0.22 \\
        2124249013097682944 & HIRESb & 6480 $\pm$ 183 & 4.21 $\pm$ 0.31 & 1.02 $\pm$ 0.14 & -2.35 $\pm$ 0.13 \\
        2097059980331156480 & HIRESb & 5521 $\pm$ 43 & 4.35 $\pm$ 0.11 & 0.57 $\pm$ 0.12 & -0.59 $\pm$ 0.03 \\
        2795884385252655232 & HIRESb & 6250 $\pm$ 307 & 4.57 $\pm$ 0.47 & 0.24 $\pm$ 1.16 & -0.66 $\pm$ 0.17 \\
        2105234402606388352 & HIRESb & 5309 $\pm$ 32 & 4.27 $\pm$ 0.08 & 0.18 $\pm$ 0.33 & -0.78 $\pm$ 0.03 \\
         & & ... & & & \\
    \end{tabular}
    \caption{Atmospheric parameters for systems which achieved successful convergence (Section \ref{ssec:sp_ew_balance}). The complete table can be downloaded online.}
    \label{tab:stellar_params}
\end{table}

\begin{table}[]
    \centering
    \begin{tabular}{c c c c c c c c c c }
       \hline
        Gaia DR3 source ID & $A$(Fe\,I) & $N$(Fe\,I) & $A$(Fe\,II) & $N$(Fe\,II) & $A$(Ba\,II) & $N$(Ba\,II) & $A$(Y\,II) & ... \\
        \hline
        \hline
        2808994137268673152 & 7.17 +/- 0.03 & 31 & 7.04 +/- 0.05 & 12 & 2.10 +/- 0.09 & 2 & 1.89 +/- 0.20 \\
        2124249013097682944 & 5.08 +/- 0.02 & 75 & 5.06 +/- 0.07 & 6 & 2.02 +/- 0.07 & 2 & 1.20 +/- 0.20 \\
        2097059980331156480 & 6.87 +/- 0.01 & 63 & 6.84 +/- 0.02 & 17 & 1.47 +/- 0.04 & 2 & 1.50 +/- 0.04 & ...\\
        2795884385252655232 & 6.79 +/- 0.02 & 20 & 6.76 +/- 0.04 & 10 & 2.97 +/- 0.09 & 2 & 3.01 +/- 0.02 \\
        2105234402606388352 & 6.68 +/- 0.01 & 60 & 6.66 +/- 0.02 & 14 & 1.38 +/- 0.10 & 2 & 1.29 +/- 0.01 \\
         & &  &  ... &  &  &  &  \\
        
    \end{tabular}
    \caption{Absolute abundances derived for systems in Table \ref{tab:stellar_params} (Section \ref{ssec:abundances_derivation}). The reported errors are the standard deviations of abundances inferred from multiple lines divided by square root of the number of lines ($N$). In the case of a single line, we adopt a fixed error of 0.2 dex. The complete table can be downloaded online.}
    \label{tab:abuns}
\end{table}

\subsection{Trends with yttrium abundance} \label{appendix:yII_trends}

\begin{figure*}
    \centering
    \includegraphics[width=0.95\linewidth]{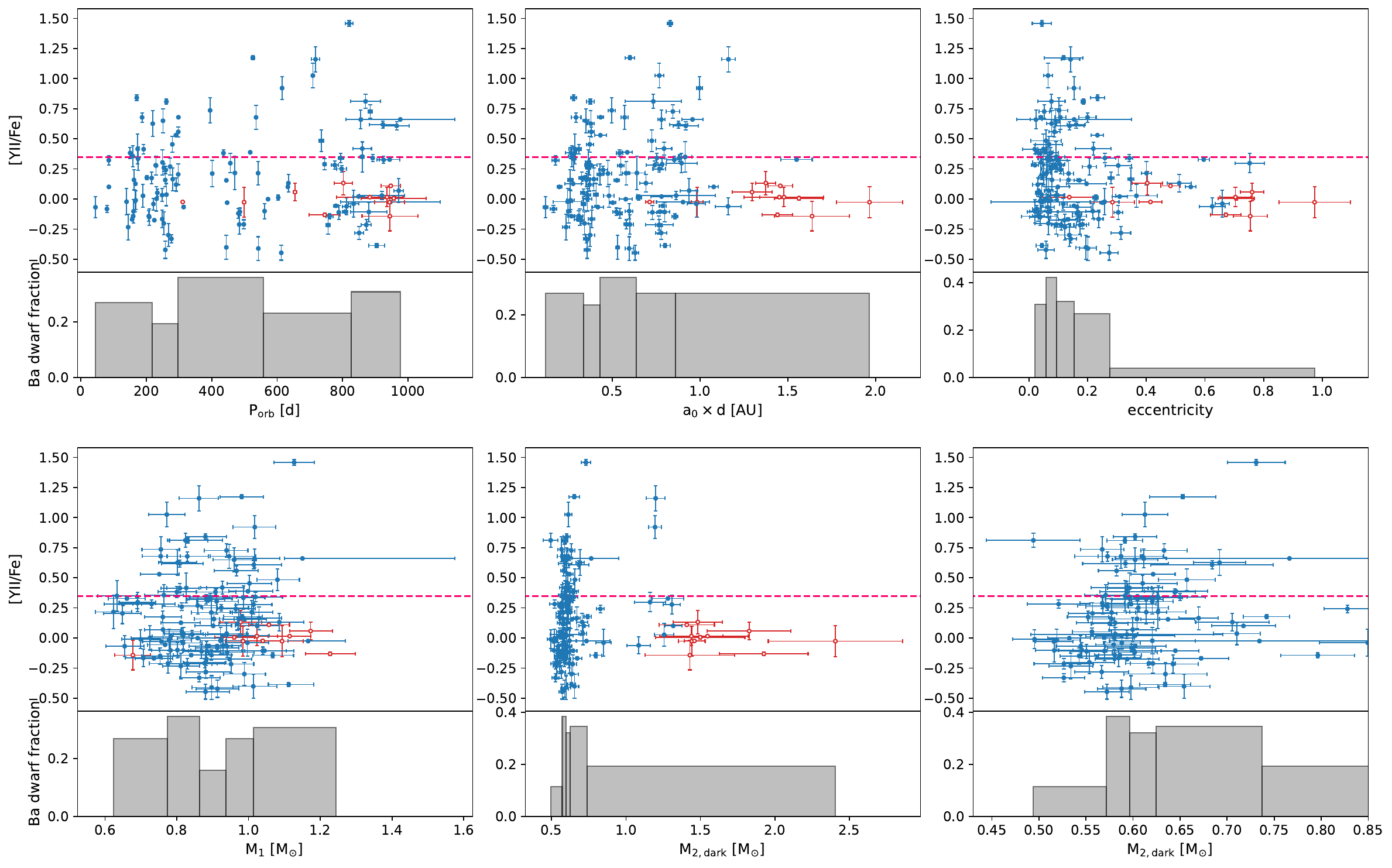}
    \caption{Analogous to Figure \ref{fig:baII_params}, but plotting [Y/Fe] instead of [Ba/Fe].}
    \label{fig:yII_params}
\end{figure*}

Figure \ref{fig:yII_params} plots [Y/Fe] as a function of several orbital and stellar parameters, analogous to that of [Ba/Fe] in Figure \ref{fig:baII_params} where similar trends (or lack thereof) are found (Section \ref{ssec:param_dependence}). 

\bibliography{refs}{}
\bibliographystyle{aasjournalv7}



\end{document}